\documentclass[12pt]{article}

\usepackage{scicite}

\usepackage{times}

\usepackage{epsfig}
\usepackage{amsmath}
\usepackage{amssymb}
\usepackage{ifthen}
\usepackage{txfonts}
\usepackage{rotating}
\usepackage{url}
\usepackage{varioref}
\usepackage{verbatim}
\usepackage{latexsym}
\usepackage{graphicx}
\usepackage{rotating}

\newcommand*{\bfrac}[2]{\genfrac{}{}{0pt}{}{#1}{#2}}
  
\DeclareSymbolFont{cmletters}{OML}{cmm}{m}{it}
  
\DeclareMathSymbol{v}{\mathord}{cmletters}{"76}

\def\be{\begin{equation}}
\def\ee{\end{equation}}

\newcommand{\msun}{{\rm M_{\odot}}}

\newcommand{\alf}{Alfv\'en}

\newcommand{\cut}[1]{\hbox{}}

\DeclareSymbolFont{cmletters}{OML}{cmm}{m}{it}
\DeclareMathSymbol{v}{\mathalpha}{cmletters}{"76}

\usepackage{subfigure}
\usepackage{ifthen}
\usepackage[usenames,dvipsnames]{color}
\usepackage{graphicx}

\usepackage{hyperref}

\newenvironment{myindentpar}[1]%
 {\begin{list}{}%
         {\setlength{\leftmargin}{#1}}%
         \item[]%
 }
 {\end{list}}

\newenvironment{sciabstract}{%
\begin{quote} \bf}
{\end{quote}}

\newcounter{lastnote}

\newcommand\mytitle{{\bf Alignment of Magnetized Accretion Disks and Relativistic Jets with Spinning Black Holes}}

\title{\mytitle}

\author
{Jonathan C. McKinney,$^{1 2\ast}$ Alexander Tchekhovskoy,$^{3}$ Roger D. Blandford$^{1}$\\
\\
\normalsize{$^{1}$Kavli Institute for Particle Astrophysics and Cosmology, Stanford University,}\\
\normalsize{PO Box 20450, MS 29, Stanford, CA 94309, USA}\\
\normalsize{$^{2}$University of Maryland at College Park, Dept. of Physics, Joint Space-Science Institute}\\
\normalsize{1117 John S. Toll Building \#082, College Park, MD 20742, USA}\\
\normalsize{$^{3}$Center for Theoretical Science, Jadwin Hall, Princeton University, Princeton,}\\
\normalsize{NJ 08544 USA; Princeton Center for Theoretical Science Fellow}\\
\\
\normalsize{$^\ast$To whom correspondence should be addressed; E-mail:  jcm@umd.edu .}
}

\date{}

\begin{document}

\baselineskip24pt

\maketitle

\begin{sciabstract}

  Accreting black holes (BHs) produce intense radiation and powerful
  relativistic jets, which are affected by the BH's spin magnitude and
  direction.  While thin disks might align with the BH spin axis via
  the Bardeen-Petterson effect, this does not apply to jet systems
  with thick disks.  We used fully three-dimensional general
  relativistic magnetohydrodynamical simulations to study accreting BHs with
  various BH spin vectors and disk thicknesses with magnetic flux
  reaching saturation.  Our simulations reveal a ``magneto-spin
  alignment'' mechanism that causes magnetized disks and jets to align
  with the BH spin near BHs and further away to reorient with the
  outer disk.  This mechanism has implications for the evolution of BH
  mass and spin, BH feedback on host galaxies, and resolved BH images
  for SgrA* and M87.

\end{sciabstract}

Astrophysical black holes (BHs) operate as engines that
convert gravitational binding energy of accreting plasmas into intense radiation \cite{sha73}
and release BH spin energy \cite{bz77,koi02,2011MNRAS.418L..79T} into powerful relativistic jets
\cite{1976Natur.262..649L,1979Natur.279..770B}.
Relativistic jets from accreting BHs
are commonly observed to emerge from active galactic nuclei (AGN) or quasars, x-ray binaries as
microquasars, and gamma-ray burst (GRB) events.  GRB jets
allow one to probe the earliest epochs of star formation,
whereas radiation and jets from AGN play a direct dynamical role
via feedback that suppresses star formation in their host galaxies \cite{dim05}.

BHs are also intrinsically interesting because they act as
laboratories for probing Einstein's general relativity theory
and for testing theories about accreting BHs and jets.
Astrophysical BHs are characterized primarily by their mass ($M$) and
dimensionless spin angular momentum ($j$).
BHs have been measured to have masses of tens to
billions of solar masses, like M87's BH with $M\sim 6\times
10^9\msun$ \cite{2011ApJ...729..119G}.
Spins have also been measured and span the whole range of
possible values, including near the maximal value of $j\sim 1$ in
the BH x-ray binary GRS1915+105 with $M\sim 14\msun$
and in the AGN MCG-6-30-15 with $M\sim 3\times 10^6\msun$ \cite{rf08,mnglps09}.
Structures on BH event horizon length scales have recently been resolved
by Earth-sized radio telescope interferometry for
SgrA* \cite{2008Natur.455...78D,2011ApJ...727L..36F}
and M87 \cite{2011Natur.477..185H,Doeleman19102012}.

The BH spin/rotational axis generally
points in a direction that is tilted by some angle $\theta_{\rm tilt}$
relative to the rotational axis of the plasma disk and its magnetic
field's direction at large distances.  While the BH's present angular
momentum axis is set by the history of plasma accretion and mergers
with other BHs, the gas being currently supplied (at mass accretion
rate $\dot{M}$) to the BH can have an arbitrarily different angular
momentum axis.  This tilt influences the intensity of the radiation via
changes in the gravitational potential felt by the plasma and also
influences what an observer at different viewing angles
sees due to disk warping and jet bending.

One mechanism known to possibly affect the orientation of a disk or
jet is the ``Bardeen-Petterson'' (BP) effect
\cite{1975ApJ...195L..65B,1981ApJ...247..677H,1985MNRAS.213..435K,2000MNRAS.315..570N},
where Lense-Thirring (LT) forces induced by the BH frame-dragging cause a
misaligned disk to precess and warp until a local viscosity
aligns a very thin disk out to some distance (estimated
to be out to $r\sim 10r_g$--$10^5r_g$,
$r_g$ is a gravitational radius\cite{note2},
depending upon assumptions) from the
BH.  Whereas the viscosity has been thought to result from
turbulence driven by the magneto-rotational instability (MRI)
that amplifies weak small-scale ($\lesssim
H$, the disk height) magnetic fields \cite{bal91},
magnetohydrodynamical (MHD) simulations of weakly magnetized disks have not yet
seen any BP alignment effect \cite{2007ApJ...668..417F}.
The BP effect and LT precession remain commonly invoked mechanisms
to understand the way tilt affects how BH mass and spin evolve
\cite{1980Natur.287..307B,2009MNRAS.399.2249P},
how merging BHs are affected by any nearby plasma
\cite{2007ApJ...661L.147B}, and how disks and their jets are oriented
\cite{1999MNRAS.309..961N,2012PhRvL.108f1302S}.

Large-scale electromagnetic (EM) fields might also affect the
jet's and disk's orientation via external confinement forces \cite{1977A&A....58..175K,mtb12}.
Estimates based upon large-scale magnetic fields being weaker than turbulent disk fields
suggested that EM forces are insufficient to align the disk
with the BH \cite{1977A&A....58..175K} or the BH with the disk
\cite{2003JCAP...09..001K,2005AIPC..784..175K}.
Simulations without disks have given ambiguous results for the EM jet direction.
For a uniform vertical magnetic field and no disk,
the jet is directed along the magnetic field direction rather than the BH's tilted spin
axis \cite{2010PhRvD..82d4045P}, whereas isolated magnetic threads
tend to align with the BH spin axis when there is no disk to restrict
their motion \cite{2004Sci...305..978S}.

We have used fully three-dimensional (3D) general relativistic (GR)
MHD simulations \cite{supp} of accreting BHs to show
that near the BH both the disk and jet reorient and align with the
BH's spin axis.  Our simulations were designed so that the magnetic
field built-up to a natural saturation strength
with, roughly, the disk's thermal+ram+gravitational
forces balancing the disk's and jet's magnetic forces
such that the trapped large-scale magnetic field threading
the BH and disk became strong compared to the disk's turbulent field.
The saturated field strength has been demonstrated
to be independent of the strength of the initial magnetic field
when the surrounding medium has a sufficient supply of magnetic flux,
weakly dependent upon BH spin, and proportionally dependent
upon disk thickness
\cite{2011MNRAS.418L..79T,mtb12,2012MNRAS.423L..55T}.
We considered various BH spins \cite{note1}, BH tilts,
and disks with a quasi-steady state height $H$ to radius $R$ ratio
of $H/R\sim 0.6$ for thick disks and $H/R\sim 0.3$ for thinner disks.
Numerical convergence of our results was determined
based upon both convergence quality measures for how well the MRI
and turbulence was resolved (Table~S2)
and based upon explicit convergence testing\cite{supp}.

Let us motivate these MHD simulations by estimating
whether EM forces are expected to dominate
LT forces on the rotating heavy disk.  Imagine a toy
model with a flat heavy disk tilted and pushed-up against
the magnetized jet generated directly by the rotating BH.
The EM torque per unit area by the jet on the disk
is $\tau_{\rm EM} \sim r B_r B_\phi/4 \sim r^2 B^2_r \Omega_{\rm F}/4$
for magnetic field $B$ bending on scale $r$
with field line rotation frequency $\Omega_{\rm F}\sim j/8$ for $j\sim 1$ \cite{mtb12}.
Introducing a dimensionless magnetic flux of
$\Upsilon\approx 0.7(4\pi r^2 B_r)/\sqrt{4\pi r_g^2 \dot{M} c}$
with $B$ in Gaussian units ($r_g$ and $c$ reintroduced for dimensional clarity)
that is consistent with measurements in our previous works \cite{mtb12},
then $\tau_{\rm EM} \sim \Upsilon^2\dot{M} \Omega_{\rm F}/(8\pi r^2)$.
Meanwhile, the LT torque per unit area is $\tau_{\rm LT} \sim \Omega_{\rm LT} L$
with LT precession rate $\Omega_{\rm LT}\sim 2j/r^3$,
disk angular momentum per unit area $L\sim \Sigma r v_\phi$,
disk surface density $\Sigma\sim \dot{M}/(2\pi r v_r$),
and far from the horizon an effective viscosity $\alpha_{\rm eff}\sim v_r/((H/R)^2 v_\phi)$.
This gives $\tau_{\rm LT}\sim j \dot{M} v_\phi/(\pi r^3 v_r)$.
The ratio of the EM to LT torques for $j\sim 1$ is then
$\tau_{\rm EM}/\tau_{\rm LT} \sim \Upsilon^2 r v_r/(64 r_g v_\phi)$,
with $r_g$ reintroduced for dimensional clarity.
Far beyond the horizon,
\begin{equation}\label{ratio}
\frac{\tau_{\rm EM}}{\tau_{\rm LT}} \sim \frac{1}{64}\Upsilon^2 \frac{r\alpha_{\rm eff}}{r_g} (H/R)^2 .
\end{equation}
Over the horizon and in the jet,
$\Upsilon\sim 10$ for our thinner disk models
and $\Upsilon\sim 17$ for our thick disk models \cite{mtb12}.
Also, for both thicknesses, $(r/r_g)\alpha_{\rm eff}\sim 15$
and roughly constant with radius
while $v_r/v_\phi\sim 1$ near the horizon \cite{mtb12}.
So, at all distances, the jet's EM forces lead to
$\tau_{\rm EM}/\tau_{\rm LT}\sim 2$ for our thinner disk models
and $\tau_{\rm EM}/\tau_{\rm LT}\gtrsim 5$ for thick disk models.
So, we expect EM forces to dominate LT forces
for both our thinner and thick disk models
(including for small spins \cite{supp}).

EM alignment forces are effective when they are larger than
forces associated with the newly accreted rotating plasma
with torque per unit area of $\tau_{\rm acc} \sim \dot{M} v_\phi/(2\pi r)$.
So, $\tau_{\rm EM}/\tau_{\rm acc} \sim \Upsilon^2 \Omega_{\rm F}/(4 r
v_\phi)$, and when these torques are equal
one obtains an implicit equation for a
``magneto-spin alignment'' radius of
\begin{equation}\label{rmsa}
r_{\rm msa}\sim \frac{\Omega_{\rm F} r_g^2\Upsilon^2}{4v_\phi} ,
\end{equation}
(with $r_g$ reintroduced),
within which EM forces can torque the accreting dense material.
For sufficiently small $\Upsilon$ or $j$, no alignment can occur.
We obtain $r_{\rm msa}\gtrsim 30r_g$
for our thinner and thick disk models that are
sub-Keplerian by a factor $0.5$ to $0.1$, respectively \cite{mtb12},
although accurate estimates require more physics \cite{supp} or simulations.

Our self-consistent fully 3D GRMHD simulations started with a disk around an untilted BH
where the BH spin axis, disk rotational axis, and emergent jet's direction all pointed in the vertical ($z$) direction.
As the simulation proceeded, the mass and magnetic flux readily advected from large distances onto the BH.
The magnetic flux vs. radius saturated on the BH and within the disk near the BH
once magnetic forces balanced the disk's thermal+ram+gravitational forces.
Magnetic braking causes such disks to become even more sub-Keplerian
than weakly magnetized thick disks \cite{mtb12},
which means the classical thin disk inner-most stable circular orbit position
is even less applicable than for weakly magnetized thick disks.
The simulations were evolved for a long time period so that the disk reached a
quasi-stationary magnetically saturated state out to about $r\sim 40r_g$ \cite{mtb12,supp}.

Then, the BH spin axis was instantly tilted by an angle of $\theta_{\rm tilt,0}$
(see Table~1 for tilts used for different spins and disk thicknesses).
The tilted disk-jet system underwent a violent rearrangement for the larger tilts.
The frame-dragging forces caused the nearly split-monopole BH magnetosphere to align with the BH spin axis,
as expected because the misaligned angular momentum
was radiated away as part of the electromagnetic outflow on~\alf~time scales.
Then, the magnetic torque caused the heavy disk to lose its misaligned component of angular momentum
and so reorient with the BH's rotating magnetosphere.
The timescale for alignment seems to be roughly the~\alf~crossing time
near the heavy disk and BH.
This ``magneto-spin alignment'' occurs because the magnetic
field built-up to a natural saturation strength on the horizon where
$\Upsilon\sim 10$ (depending upon the thickness and tilt), which led
to the BH magnetosphere's forces dominating the disk dynamics and LT forces.

All simulations (including with zero tilt) were then further evolved in time
until all the tilted simulations reached a
new quasi-stationary state out to $r\sim 40r_g$.  This ensured that
any differences at later times in the disk and jet between tilted and untilted
simulations were due to the BH tilt. This also ensured
that each of the tilted and untilted simulations reached their own quasi-steady state values for the
magnetic flux near the BH, magnetic flux in the disk, mass accretion rate, etc.
We then measured the evolved relative tilt between the BH spin axis and the
disk \& jet at $r=4r_g$ and $r=30r_g$ (Table~1).
For all our models, the disk \& jet aligned with the BH spin axis near the BH,
whereas the disk axis \& jet direction deviated at larger distances.
Such deviations are expected because the jet interacted
with circulation with stronger mass inflows at larger distances\cite{supp}.
Despite the tilts and jet deviations, the BH's efficiency (defined as the
ratio of energy out to energy in) was roughly $100\%$ for $j\gtrsim 0.9$
(Table~S1),
where more tilt led to reduced efficiency
due to more spatially and temporally irregular mass inflow.

\begin{table}
\caption{Tilted Black Hole Disk-Jet Systems}
\begin{center}
\begin{tabular}[h]{|l|r|r|r|r|r|r|r|}
\hline
ModelName         &        BH       $j$    &   Disk $H/R$ &    Initial Tilt    &  $\bfrac{\rm Disk Tilt}{r=4r_g}$   &  $\bfrac{\rm Jet Tilt}{r=4r_g}$    &   $\bfrac{\rm Disk Tilt}{r=30r_g}$  &    $\bfrac{\rm Jet Tilt}{r=30r_g}$  \\
\hline
A0.94BfN40        &        0.9375          &      0.6        &        0            &          0            &           0           &             0          &            0           \\
A0.94BfN40T0.35   &        0.9375          &      0.6        &        0.35         &          0.0          &           0.0         &             0.2        &            0.2         \\
A0.94BfN40T0.7    &        0.9375          &      0.6        &        0.70         &          0.0          &           0.0         &             0.4        &            0.3         \\
A0.94BfN40T1.5708 &        0.9375          &      0.6        &        1.5708       &          0.0          &           0.1         &             0.5        &            0.7         \\
\hline
A-0.9N100         &        -0.9            &      0.3        &        0            &          0            &           0           &             0          &            0           \\
A-0.9N100T0.15    &        -0.9            &      0.3        &        0.15         &          0.1          &           0.1         &             0.1        &            0.2         \\
A-0.9N100T0.3     &        -0.9            &      0.3        &        0.30         &          0.2          &           0.2         &             0.2        &            0.2         \\
A-0.9N100T0.6     &        -0.9            &      0.3        &        0.60         &          0.2          &           0.3         &             0.4        &            0.3         \\
A-0.9N100T1.5708  &        -0.9            &      0.3        &        1.5708       &          0.2          &           0.4         &             0.9        &            0.8         \\
\hline
A0.9N100          &        0.9             &      0.3        &        0            &          0            &           0           &             0          &            0           \\
A0.9N100T0.15     &        0.9             &      0.3        &        0.15         &          0.0          &           0.0         &             0.1        &            0.1         \\
A0.9N100T0.3      &        0.9             &      0.3        &        0.30         &          0.1          &           0.1         &             0.2        &            0.2         \\
A0.9N100T0.6      &        0.9             &      0.3        &        0.60         &          0.1          &           0.1         &             0.3        &            0.3         \\
A0.9N100T1.5708   &        0.9             &      0.3        &        1.5708       &          0.2          &           0.3         &             0.7        &            0.6         \\
\hline
A0.9N100          &        0.99            &      0.3        &        0            &          0            &           0           &             0          &            0           \\
A0.99N100T0.15    &        0.99            &      0.3        &        0.15         &          0.0          &           0.1         &             0.1        &            0.1         \\
A0.99N100T0.3     &        0.99            &      0.3        &        0.30         &          0.1          &           0.1         &             0.2        &            0.2         \\
A0.99N100T0.6     &        0.99            &      0.3        &        0.60         &          0.1          &           0.1         &             0.3        &            0.4         \\
A0.99N100T1.5708  &        0.99            &      0.3        &        1.5708       &          0.1          &           0.1         &             0.6        &            0.6         \\
\hline
\hline
\end{tabular}
\end{center}
\label{tbl0}
\caption{\addtocounter{table}{-1}The simulation models listed by model name,
  which identifies the approximate BH spin of $j=x$ (giving Ax),
  something about the magnetic field choices for some y and z (giving ByNz) \cite{mtb12},
  and the initial relative tilt ($\theta_{\rm tilt,0}=p$ in radians) between the BH spin axis
  and the disk rotation axis (giving Tp, where no T means $\theta_{\rm tilt,0}=0$).
  The second column gives the dimensionless BH spin ($j$), the third column gives the evolved
  quasi-steady state value of the disk height-to-radius ratio $H/R$ between $r\sim 20r_g$ and $r\sim 30r_g$,
  while the fourth column identifies the initial relative tilt ($\theta_{\rm tilt,0}$)
  between the disk+jet (having the same tilt initially) and the BH spin axis.
  The fifth and sixth columns give the evolved relative tilt between the BH spin axis
  and the disk \& jet, respectively, at $r=4r_g$.
  The seventh and eighth columns give the same measurements at $r=30r_g$.
  A relative tilt of $0.0$ means the disk or jet remained aligned with the BH spin axis,
  while a tilt equal to the initial tilt means the disk or jet was unaffected.}
\end{table}

\begin{figure*}
\centering
\includegraphics[width=6.6in,clip]{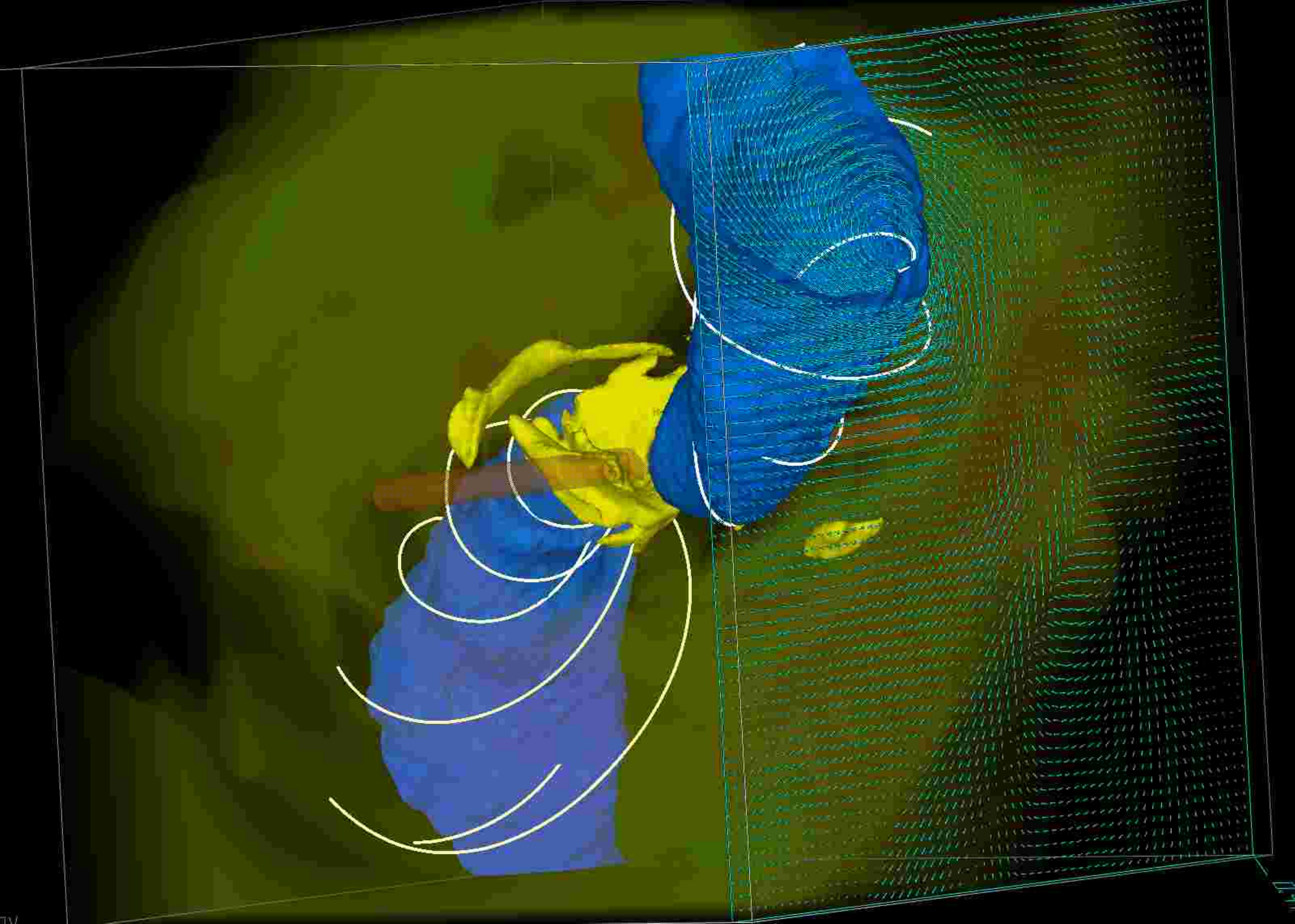}
\caption{3D snapshot for an evolved model with $j=0.99$, initial relative
  tilt $\theta_{\rm tilt,0}\approx90^\circ$, and disk thickness of $H/R\sim 0.3$.  The
  rotating BH sits at the center of the box of size $r=-40r_g$ to $r=+40r_g$
  in each dimension.  The snapshot shows the disk near the BH (yellow
  isosurface, which is mostly flat in the figure plane), the highly
  magnetized jet region (blue isosurface, with magnetic energy per
  unit rest-mass energy equal to about $70$), the rotational axis of
  the disk both initially and at large distances (orange cylinder),
  outer disk (green-yellow volume rendering, more aligned with disk rotational axis at large distances),
  magnetic field vectors (like iron filings on
  that surface) for a cross-section of the jet (cyan vectors), and jet
  magnetic field lines (white lines) that trace from the BH out to
  large distances.  The disk and jet near the BH are aligned with the
  BH spin axis and point mostly in-out of the figure plane, while at
  larger distances the jet points roughly half-way between the BH spin
  axis and the disk's rotational axis at large distances (pointing
  along the orange cylinder).}
\label{fullrotvec4}
\end{figure*}

\begin{figure*}
\centering
\includegraphics[width=6.6in,clip]{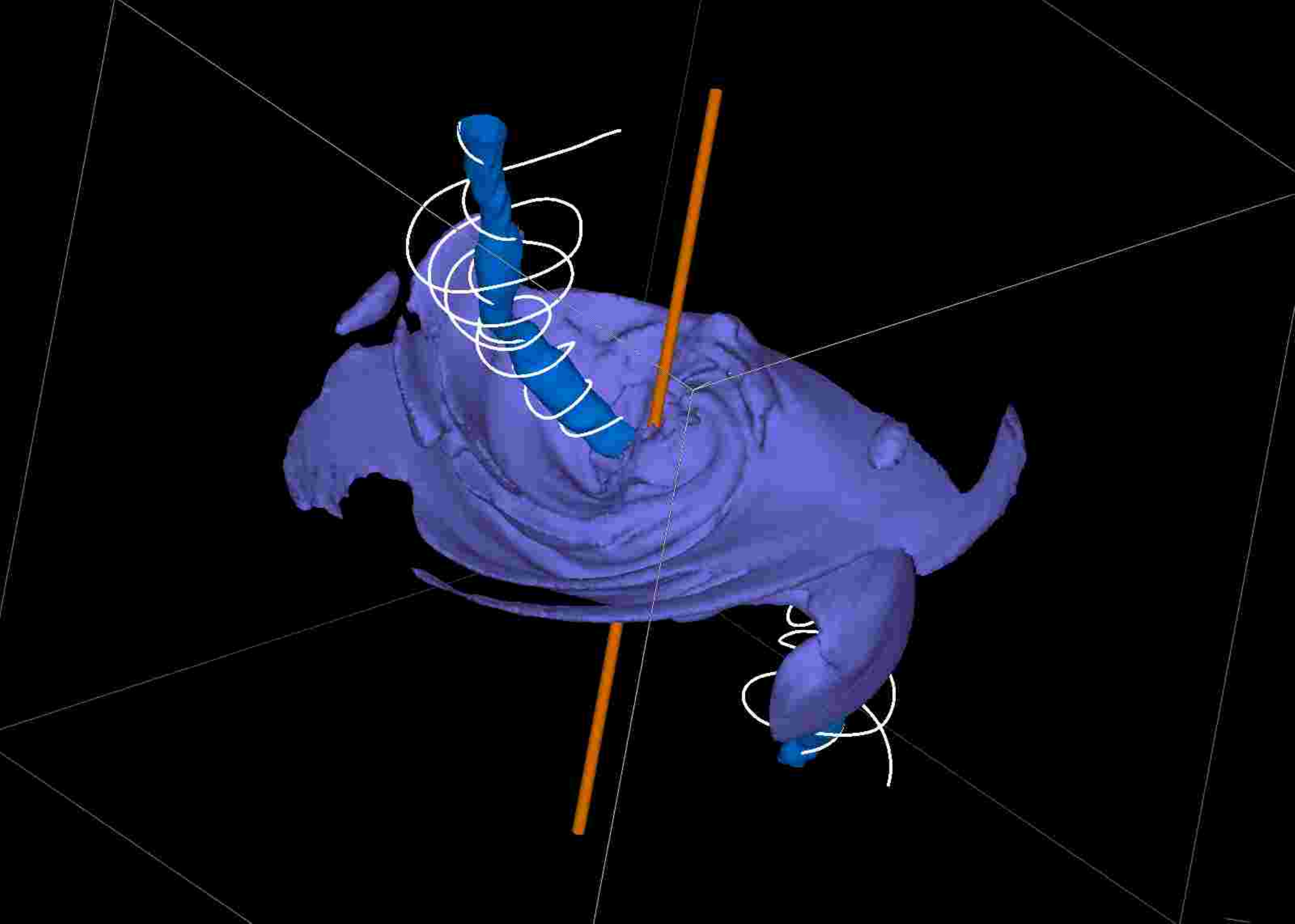}
\caption{3D snapshot that is similar to Fig.~\ref{fullrotvec4} but
  shows the outer disk (density isosurface in purple) at large
  distances from the BH in a box of size $r=-350r_g$ to $r=350r_g$ in each
  dimension.  The jet (blue isosurface, here with magnetic energy per
  unit rest-mass energy equal to about $4$, still corresponding to the jet spine)
  is aligned with the BH spin near the BH but gradually gets pushed by the disk material and
  becomes parallel to (but offset from) the disk rotational axis at
  large distances.  The strong interaction between the jet and disk has left an
  asymmetry or warp in the disk density at large radii.}
\label{largejet}
\end{figure*}

Our most extreme case of a tilted BH accretion disk and jet
system is the $j=0.99$ model with a full tilt of $\theta_{\rm
  tilt,0}=1.5708\approx 90^\circ$ and disk thickness $H/R\sim 0.3$.
Even in this extremely tilted
case, the evolved disk and jet near the BH aligned with the BH spin axis
(Fig.~\ref{fullrotvec4}, Fig.~S1, Movie~S1).
The jet's magnetic field winded around the persistent relativistic jet,
and the magnetic field was
well-ordered even for this highly tilted case.  The jet was not
symmetric around the jet axis, and instead there was a broad wing (with
opening half-angle of about $25^\circ$ by $r=40r_g$) and a narrow wing
(with opening half-angle of about $5^\circ$ by $r=40r_g$).
At large distances from the BH,
the jet drilled its
way through the disk material and gradually got
pushed away from the disk (Fig.~\ref{largejet} and Movie~S2).
The magnetic field winded around with a pitch angle of about
$45^\circ$ near the BH and had a smaller pitch angle at larger distances.
By $r\sim 300r_g$, the jet had become
parallel with (but offset from) the outer disk rotational axis,
and so the jet and counter-jet were also offset.

Thus, our simulations have revealed a ``magneto-spin alignment''
mechanism that aligns the disk \& jet axes with the BH spin axis near
the BH once the magnetic field has saturated on the BH and within the
disk \cite{note3}.
Unlike the BP effect, the mechanism actually works best for thick disks,
and so the magneto-spin alignment mechanism should control jet systems where
thick disks (due to accretion at either very low\cite{1995Natur.374..623N}
or very high rates when $H/R\gtrsim 0.5$) are expected.
For example, for SgrA* and M87, if the BHs rotate sufficiently rapidly \cite{supp},
then we expect their jet's and disk's
photon spectra,
temporal behaviors,
and resolved images
to be affected by
non-zero relative tilts due to disk warping and jet bending near the BH.

Tidal disruption flare events like Swift J164449.3 +573451 are thought
to be produced by very high accretion rates onto BHs,
which launch fairly persistent jets that dissipate and give the observed emission
\cite{2011Sci...333..203B}.
Our results suggest the inner disk and inner jet are both aligned with the BH
spin axis, but the observed jet dissipating at large distances
need not point along the BH spin axis.
EM forces do not directly cause any precession \cite{1977A&A....58..175K},
so the lack of LT precession-induced variability does not alone
imply that the jet is necessarily driven by the BH spin power \cite{2012PhRvL.108f1302S}.
Further, quasi-periodic oscillations
\cite{Reis24082012} and long-term dips seen in
this system's light curve might be explained by oscillations in
the disk-jet magnetospheric interface \cite{mtb12}
or by periods of magnetic flux accumulation and rejection
by the BH \cite{2011MNRAS.418L..79T} (both occurring for untilted
systems) rather than by LT precession.

Jet dissipation/emission (e.g. in blazars) might be due to the jet
ramming into the disk until the jet aligns with the disk rotational
axis at large distances.  Measurements of BH spin in AGN and x-ray
binaries might be affected by assumptions about the alignment
between the disk, BH spin, and jet \cite{rf08,mnglps09}.  The
cosmological evolution of BH mass and spin and AGN feedback for
accretion at high rates might be affected by the higher BH
spin-down rates and jet efficiencies compared to standard thin disk
spin-down rates and radiative efficiencies \cite{mtb12} and also by how the
jet aligns the disk material before LT torques can be effective so
possibly leading to less change in the BH spin direction compared to
the BP effect.

\bibliography{bib}

\begin{thebibliography}{10}
\bibitem{note2}
We typically set $GM=c=1$, where $c$ is the speed of light, $G$ is the
  gravitational constant, and $M$ is the mass of the BH, so that $r_g\equiv
  GM/c^2=1$. For dimensional clarity, these constants are sometimes
  reintroduced.
\bibitem{note3}
The magnetic field built-up via direct magnetic flux advection, but the
  build-up might also occur via dynamo generation as seen in our prior untilted
  simulations that showed emergent large-scale dipolar flux patches.
\end{thebibliography}

\bibliographystyle{Science}

\paragraph*{Acknowledgements:}
\begin{myindentpar}{1cm}
  JCM thanks Narayan, Dexter, and Fragile for useful discussions, and
  Ralf Kaehler at KIPAC (SLAC/Stanford) for the artistic rendering in
  Figure S1.  This work was supported by a NASA Fermi grant NNX11AO21G
  (JCM), Princeton Center for Theoretical Science Fellowship (AT) and
  NSF/XSEDE resources provided by TACC (Lonestar/Ranch) and NICS
  (Kraken) under the awards TG-PHY120005 (JCM) and TG-AST100040 (AT)
  and provided by NASA (NAS Pleiades) for the Fermi grant.
  GRMHD simulation data are contained in Table~1 and Tables~S1 to~S2
  in the Supporting Online Materials.

\end{myindentpar}

\paragraph*{Supplementary Materials:}
\begin{myindentpar}{1cm}
www.sciencemag.org\\
Materials and Methods\\
Fig. S1 \\
Tables S1 S2 \\
Movies S1 S2 \\
References (40-98) \\ 
\end{myindentpar}

\end{document}


\baselineskip24pt

\begin{figure*}
\centering
\includegraphics[width=1.63in,clip]{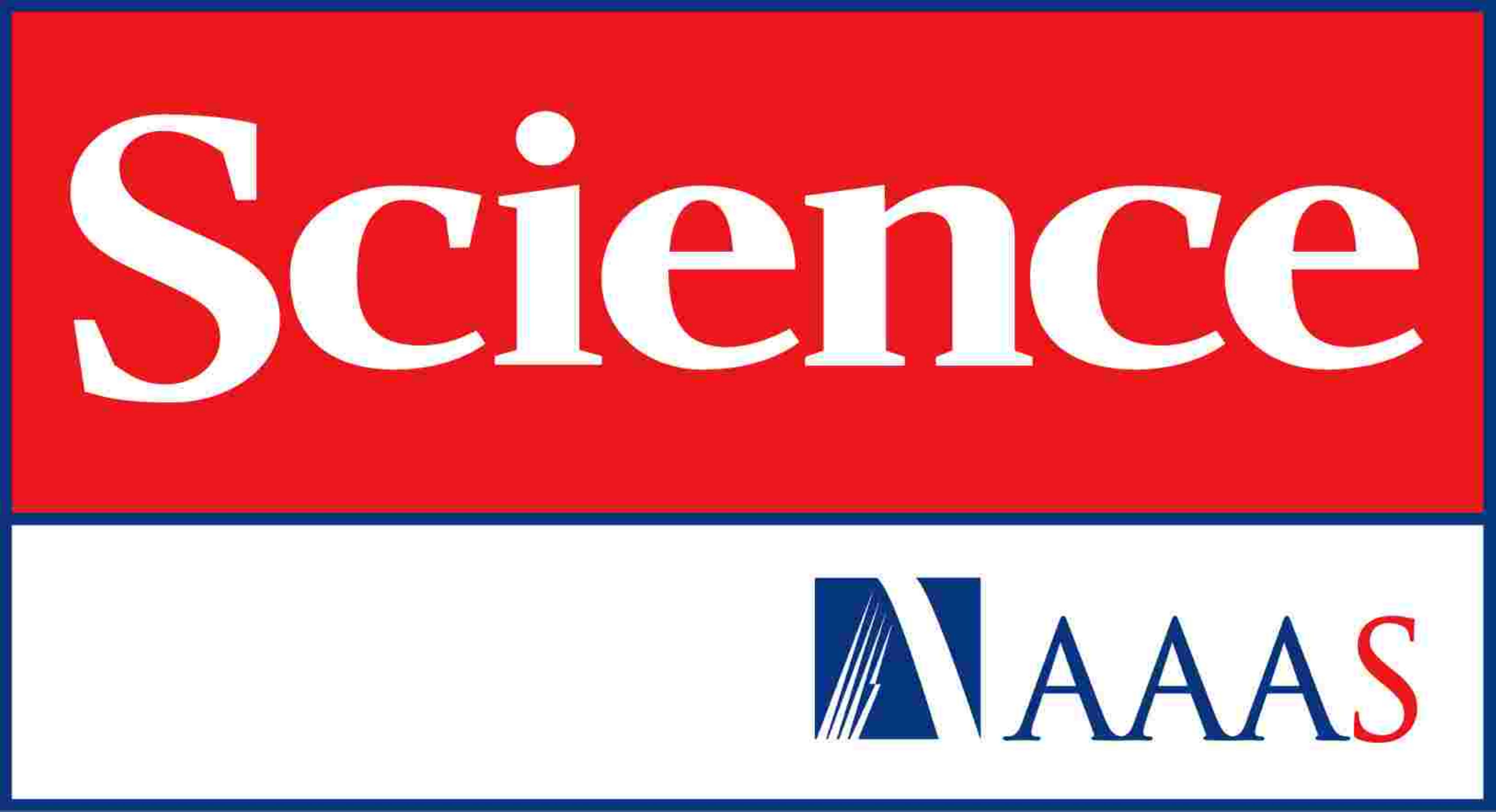}
\end{figure*}

\maketitle

\begin{sciabstract}

\end{sciabstract}

{\bf This PDF file includes:}

{\indent\ \ \ \ \  Materials and Methods}\\
{\indent\ \ \ \ \  Fig. S1}\\
{\indent\ \ \ \ \  Tables S1 to S2}\\
{\indent\ \ \ \ \  Captions for Movies S1 to S2}\\
{\indent\ \ \ \ \  References (40-98)}

{\bf Other Supplementary Materials for this manuscript includes the following:}

{\indent\ \ \ \ \  Movies S1 to S2}

\newpage

\section{Methods}

In this supporting text, we describe our numerical methods, elaborate
on details for the toy model in the report that estimates the effects
of magnetic fields vs. Lense-Thirring precession around tilted black
holes (BHs), present the equations of motion we numerically solve,
discuss the diagnostics computed, and summarize the setup of our
physical and numerical models.

\subsection{Numerical Methods}
\label{sec:methods}

We used fully three-dimensional (3D) general relativistic
magnetohydrodynamic (GRMHD) simulations to study BH
accretion flows with various BH spins (dimensionless value of $j$),
relative tilts between the BH spin axis and the angular momentum
(rotational) axis at large radii ($\theta_{\rm tilt,0}$ in radians), and
both thick and thinner disks (height $H$ to radius $R$ ratio of $H/R\sim
0.6$ and $H/R\sim 0.3$, respectively).

Our simulations used the GRMHD code HARM based upon a conservative
shock-capturing Godunov scheme with 2nd order Runge-Kutta
time-stepping, Courant factor $0.8$, Lax Friedrichs (LAXF) fluxes,
simplified wave speeds, piecewise parabolic method (PPM)
type interpolation (with no contact steepener, but with
shock flattener based upon rest-mass flux density and total pressure),
a staggered magnetic field representation, and a regular grid warping
\citep{gam03,nob06,tch_wham07} as used in our prior works(28).
A free version of HARM is available online for download
\cite{2012ascl.soft09005G}.
The HARM code has several key features that make it suitable for highly
magnetized regions around BHs with tilted orientations (see
appendix of (28)).  In particular, the spherical polar
coordinate axis is treated using transmissive (not reflective)
boundary conditions so that the dissipation near the axis is kept to a
minimum.  This transmissive boundary condition treats all quantities
in boundary cells near the axis as smooth and continuous functions
whose values coincide with the non-boundary cells on the other side of
the axis.  This allows the flow and any components of the magnetic
field to pass directly through the polar axis with little artificial
dissipation while maintaining a robust solution via the Godunov
scheme.

\subsection{Toy Model}
\label{sec:tilteffect}

In this section, we give more details about the toy model
introduced in the report.

The nature of jets around BHs has been a topic of interest for
a long time \cite{1966Natur.211..468R,1968Natur.219..127R,1977Natur.267..211B,1976Natur.260..683F,1978Natur.275..516R,1982Natur.295...17R,2012ApJ...757L..14L}.
Our focus is on how the disk and jet are affected by a relative tilt
with respect to the BH spin axis.
Tilted BHs have been studied at an idealized level for a few decades.
Solutions have been found for static scalar and vacuum magnetic fields around
rotating BHs with relative tilt \cite{1972ApJ...175..243P,1976RSPSA.350..239P,1977RSPSA.356..351P,1985MNRAS.212..899B,1989SvPhU..32...75A}.
For a vacuum magnetic field, it was found that the BH would be
penetrated by non-zero magnetic flux for a maximally spinning
($j=1$) BH only if there were relative tilt \cite{1985MNRAS.212..899B}.  Torque on a
charged BH was also considered \cite{1987Ap&SS.135...81A}.
For a review of these idealized works, see \cite{1989SvPhU..32...75A}.
Simulations have been required in order to push beyond these early idealized models,
with some advances for simulations without tilt (for a review, see \cite{2002Sci...295.1653B,2005Sci...307...77N}).
Some axisymmetric simulations without tilt also see flow reversal near the BH \cite{2002ApJ...577..866D},
with the effect being quite strong once the magnetic field strength saturates \citemtb.
Untilted simulations have so far only studied how strong magnetic fields affect
the aligned component of the BH spin \cite{gammie_bh_spin_evolution_2004} and \citemtb.
So far, no GRMHD simulations with tilt show any Bardeen-Petterson type alignment effect \citefragile.
Modern simulations of relatively thin weakly magnetized disks tilted relative
to the BH spin axis have been studied,
including the effect of tilt on the inner-most stable circular orbit (ISCO) \cite{2009ApJ...706L.246F},
the development of shocks in the disk near the BH \cite{2008ApJ...687..757F},
the effect of tilt on observations of BH accretion disks \cite{2011ApJ...730...36D,2012arXiv1204.4454D},
and the effect on neutron star-BH mergers \cite{2012arXiv1209.1632E}.
Simulations of tilted disks with naturally saturated magnetic field strengths
are required to understand how disks and jets are oriented under generic circumstances.
Astrophysical applications justifiably only so-far invoke the Bardeen-Petterson effect
as a possible alignment effect
\cite{1978Natur.275..516R,1998ApJ...506L..97N,2000ApJ...537..152K,2001ApJ...553..955F,2002MNRAS.330..609D,2002MNRAS.336.1371M,2005MNRAS.363...49K,2006ApJ...638..120C,2006ApJ...653..112C,2007MNRAS.379..135C,2007ApJ...667..704V,2008MNRAS.391L..15M,2010ApJ...717L..37H,2010ApJ...719L..79F,2010ApJ...713L..74F,2011MNRAS.414.1183K,2012arXiv1202.4231L},
because no other type of alignment effect has yet been demonstrated in any GRMHD simulations
and all theoretical arguments suggested that magnetic torque alignment would not occur \citeking.

Here we estimate the ratio of the electromagnetic (EM) force on the heavy disk relative
to the Lense-Thirring (LT) force on the rotating heavy disk.
This argument was used to get roughly correct the factors of order unity that appear in the report.
We also account for some other subtle effects, like disk thinning, not accounted for in the report.
Imagine a toy picture with a flat heavy disk tilted with respect to a rotating BH,
with both the BH and disk threaded by a magnetic field ($\mathbf{B}$) that bends on
scale of radius $r$.  The magnetic field is nearly co-rotating with
the BH-disk system, which drives a current ($\mathbf{J}$) into the ambient medium.  Then, the
EM torque per unit area on the disk by magnetic stresses is $\tau_{\rm EM} \equiv L_{\rm
  EM}/dt \sim |\int d\theta\sin\theta \mathbf{r} \times
(\mathbf{J}\times\mathbf{B})|\sim \pi \sin\theta_{\rm tilt} r B_r
B_\phi/(4\pi)$ for radial field $B_r$, toroidal
field $B_\phi\sim r\sin\theta\Omega_{\rm F} B_r$ for field line angular frequency of rotation of $\Omega_{\rm F}$,
and absolute magnetic flux on a radial shell of $\Phi\sim 4\pi r^2 |B_r|$
with dimensionless magnetic flux of
$\Upsilon\approx 0.7\Phi/\sqrt{4\pi r_g^2 \dot{M} c}$
for $\Phi$ in Gaussian units \citemtb.
We focus on the jet with a roughly constant magnetic flux $\Phi$,
while the magnetic field threading the disk would enhance the EM forces.
The LT torque per unit area is $\tau_{\rm LT} \equiv dL_{\rm LT}/dt
\sim \left|\mathbf{\Omega_{\rm LT}}\times \mathbf{L}\right|\sim \sin\theta_{\rm tilt}\Omega_{\rm LT}
L$ with $|\mathbf{\Omega}_{\rm LT}|=2|\mathbf{J}_{\rm
  BH}|/r^3=2j/r^3$ for disk angular momentum per unit area $L\sim
\Sigma r v_\phi$, disk surface density $\Sigma\sim \dot{M}/(2\pi r
v_r$), effective viscosity $\alpha_{\rm eff}\sim v_r/(G (H/R)^2
v_\phi)$, and $G\sim 1$ accounts for GR corrections \citemtb.
Even untilted disks that reach magnetic flux saturation undergo
magnetic braking that (in addition to the disk being thick)
leads to $\alpha_{\rm eff}\sim 1$ and quite sub-Keplerian rotation \citemtb,
which makes us expect a) smaller LT forces ; b) EM forces can more easily change the disk rotational axis ;
and c) the classical ISCO position
is even less applicable than for weakly magnetized thick disks.
The ratio of the EM to LT torques is then
$\tau_{\rm EM}/\tau_{\rm LT} \sim \Upsilon^2~(r_g\Omega_{\rm F})/(cj)~(r v_r)/(8 r_g v_\phi)$,
with $c$ and $r_g$ reintroduced for clarity.
Far from the horizon, one can write this ratio as
\begin{equation}
\frac{\tau_{\rm EM}}{\tau_{\rm LT}} \sim \frac{1}{8}\Upsilon^2 \left(\frac{r_g\Omega_{\rm F}}{cj}\right) \frac{r\alpha_{\rm eff}}{r_g} G (H/R)^2 .
\end{equation}
Assume the free parameters are
$j\sim 1$, $\dot{M}(r)$ constant as valid for $r\lesssim 30r_g$, $\Omega_{\rm F}\sim \Omega_H/4 =
j/(8(1+\sqrt(1-j^2)))\sim j/8$ for the jet magnetosphere near the disk, and $\Upsilon\sim 10$--$17$
for our thinner and thick models, respectively \citemtb.
Also, over $r\sim 12r_g$--$20r_g$, the radially averaged $\alpha_{\rm eff}\sim 1$--$0.4$
(with $(r/r_g)\alpha_{\rm eff}$ roughly constant over the solution)
and $H/R\sim 0.3$--$0.4$, respectively for the thinner and thick models,
where the thicker disk undergoes some compression by magnetic forces \citemtb.
For our thinner models we get $\tau_{\rm EM}/\tau_{\rm LT}\sim 2$ while our
thicker models get $\tau_{\rm EM}/\tau_{\rm LT}\sim 4$.
So, we expect EM forces to dominate LT forces for roughly $H/R\gtrsim 0.2$ at $r\sim 10r_g$--$20r_g$.
In addition, near the horizon, $v_r/v_\phi\sim 1$ \citemtb,
so near the horizon one obtains $\tau_{\rm EM}/\tau_{\rm LT}\sim 2$ for thinner models
and $\tau_{\rm EM}/\tau_{\rm LT}\sim 5$ for thicker models.
For non-zero spin, the value of $\Upsilon$ is roughly constant as a
function of spin for both the thinner and thick disk models \citemtb,
so these estimates suggest EM torques dominate LT torques for all spins.

The additional EM force by frame-dragging on the magnetic field threading the disk
gives $\Upsilon\propto r$ in the disk \citeretro and naively $\Omega_{\rm F}\propto r^{-3}$,
which apparently leads to LT forces eventually dominating over the disk's {\rm internal} EM forces at large radii (but the jet's EM forces external to the disk can still dominate LT forces as discussed above).
However, significant BH spin energy and angular momentum is fed directly into the disk itself \citemtb,
which leads to angular momentum transport that forces the disk's magnetic field to rotate
a bit closer to the BH magnetosphere's value of $\Omega_{\rm F}$ even far from the BH
(e.g. the rotation of the fluid flow completely re-aligns
with the BH spin direction out to $r\sim 40r_g$ for untilted retrograde
($j=-0.9375$) thick disks ($H/R\sim 0.6$) such that the flow is no longer retrograde
and has the same radial profile
of $v_\phi$ but simply with an opposite sign compared to the
otherwise identical model with $j=+0.9375$ \citemtb).

Using a similar analysis as above we can {\it roughly} estimate
whether the jet's EM forces actually affect the flow.  Such an analysis is
easier than for the BP effect that requires a model of the assumed
local viscosity that would be due to MHD turbulence, while large-scale
EM fields directly affect the plasma.  The torque per unit area
associated with the newly accreted rotating plasma at some fixed
Eulerian radius is $\tau_{\rm acc} \sim \dot{M} v_\phi/(2\pi r)$ so
that $\tau_{\rm EM}/\tau_{\rm acc} \sim \Upsilon^2 \Omega_{\rm F}/(4 r
v_\phi)$.  This gives a ``magneto-spin alignment'' radius of
\begin{equation}\label{rmsa}
r_{\rm msa}\sim \frac{\Omega_{\rm F} r_g^2\Upsilon^2}{4v_\phi}
\end{equation}
(with $r_g$ reintroduced) where the torques are equal,
within which EM forces can torque the accreting dense material.  For
our thinner disk models and $j\sim 1$ with $v_\phi\propto r^{-1/2}$
(but sub-Keplerian in magnitude by a factor of $0.5$) \citemtb, the
jet's external EM forces control the disk orientation out to $r\sim 40r_g$.
For our thick disk models and $j\sim 1$ with $v_\phi\propto r^{-1}$ (or
Keplerian with a reduction factor of $0.1$) out to $r\sim 30r_g$ up to
where a steady-state was reached \citemtb, EM forces can easily
torque the disk out to $r\sim 30r_g$
and may even dominate out to $r\sim 10000r_g$
if the flow remains the same factor below Keplerian at such large
radii (as is possible for thick disks that reach magnetic field strength
saturation out to large radii).

The magneto-spin alignment radius might depend upon spin.  It is
possible that for sufficiently slowly spinning BHs that produce weak
jets, there might be no magneto-spin alignment effect.  While rapidly
rotating BHs might exist in SgrA* and M87 as some models/estimates
show \cite{2009ApJ...706..497M} and \citedoel, this is still in
debate \cite{spm10,broder11}, so it is important to
roughly estimate how spin might affect our results.  For example, from
Equation~(\ref{rmsa}) and for the near-horizon region (i.e. $r\sim
r_g$), horizon-scale alignment due to the jet alone occurs for
$j\gtrsim 0.5$ for our thinner disk models and $j\gtrsim 0.2$ for our
thick disk models if we use $v_\phi\sim 0.5c$ as valid for the
near-horizon when $j\sim 1$ (e.g., see figure 7 in \citemtb).
However, more generally, over a range of spins, $v_\phi\sim r_g
\Omega_{\rm F}$ near the horizon due to the BH dragging the magnetic field
that then drags the plasma \citemtbandsasha, so our
estimates imply that all spins would allow for magneto-spin alignment
of the disk close to the BH event horizon.

The magneto-spin alignment radius might also depend upon the existence
of mass inflow-outflow circulation that is commonly seen in thick disk
simulations.  The BH jet value of $\Upsilon$ is determined by force
balance with the ingoing accretion near the horizon, but at larger
radii one expects thick disks to produce winds that scale as
$\dot{M}_{\rm wind}\propto r$ and so have an ingoing accretion rate
following $\dot{M}_{\rm in}\propto r$ \citemtb, such that the EM
forces are eventually ineffective against the heavy mass inflow
forces.  Stated another way, the local $\Upsilon$ becomes too small in
terms of the jet magnetic flux per unit square root of the local mass
inflow rate.  So, at some radius one expects the disk inflow to
dominate the jet and for both the disk and jet to be parallel to the
rotation axis of the outer disk (as consistent with the tilted
simulations presented in this report).

These estimates suggest that if an astrophysical system were to have a
truncated thick disk inside $r\sim r_{\rm msa}$ and still reach
magnetic field strength saturation, then the entire disk and jet would
mostly align with BH spin.

These estimates also suggest that prior GRMHD simulations of tilted
disks see no alignment effect \citefragile because their
models have $H/R\sim 0.1$--$0.2$ and have a limited amount of magnetic
flux available for accretion, both implying a small $\Upsilon$ so the
EM jet cannot dominate the disk dynamics and EM forces cannot dominate
LT forces.  For example, the standard ``thick'' torus with $H/R\sim
0.2$ with a single poloidal field loop in the initial torus is used in
many GRMHD simulations.  This setup is similar to the one we studied a
few years ago\cite{mb09} that we recently realized leads to limited
magnetic flux with $\Upsilon\sim 2$ near the BH and so leads to low
jet efficiencies despite a spin of $j\sim 0.9$ \citemtb.  Using
$\Upsilon\sim 2$ in our above analysis shows that magneto-spin
alignment does not occur for any normal range of spins (i.e. $-1\le
j\le +1$), which is consistent with existing tilted MHD
simulations \citefragile.

Given the many non-trivial effects involved, simulations are optimal
in order to obtain an accurate self-consistent solution once the
magnetic field strength saturates so that all competing forces and
effects can be accounted for.

\subsection{Governing Equations}
\label{sec:goveqns}

We solved the GRMHD equations for a magnetized accretion flow around a
tilted rotating BH defined by the Kerr metric in Kerr-Schild
coordinates with internal coordinates $\xvec^\alpha\equiv
(t,\xvec^{(1)},\xvec^{(2)},\xvec^{(3)})$ mapped to the spherical polar
coordinates $\rvec^\alpha\equiv (t,r,\theta,\phi)$.  The z-axis of the
spherical polar coordinate Kerr solution was rotated around the y-axis
by $\theta_{\rm tilt,0}$ \cite{2005ApJ...623..347F}.  We write
orthonormal vectors as $u_i$, contravariant (covariant) vectors as
$\uvec^i$ ($\uvec_i$), and higher-ranked coordinate basis tensors with
no underbar.  We work with Heaviside-Lorentz units, often set
$c=GM=1$, and let the horizon radius be $r_{\rm H}$.

Mass conservation gives
\begin{equation}
\nabla_\mu (\rho_0 \uvec^\mu) = 0 ,
\end{equation}
where $\rho_0$ is the rest-mass density, $\uvec^\mu$ is the
contravariant 4-velocity, and $\rho=\rho_0 \uvec^t$ is the lab-frame
mass density.

Energy-momentum conservation gives
\begin{equation}\label{emomeq}
\nabla_\mu  T^\mu_\nu = 0 ,
\end{equation}
where the stress energy tensor $T^\mu_\nu$ includes both matter (MA)
and electromagnetic (EM) terms:
\begin{eqnarray}\label{MAEM}
{T^{\rm MA}}^\mu_\nu &=& (\rho_0 + e_g + p_g ) \uvec^\mu \uvec_\nu + p_g \delta^\mu_\nu \nonumber ,\\
{T^{\rm EM}}^\mu_\nu &=& b^2 \uvec^\mu \uvec_\nu + p_b\delta^\mu_\nu - \bvec^\mu \bvec_\nu \nonumber ,\\
T^\mu_\nu &=& {T^{\rm MA}}^\mu_\nu + {T^{\rm EM}}^\mu_\nu .
\end{eqnarray}
The MA term can be decomposed into a particle (PA) term: ${T^{\rm
    PA}}^\mu_\nu = \rho_0 \uvec_\nu \uvec^\mu$ and an enthalpy (EN)
term.  The MA term can be reduced to a free thermo-kinetic energy
(MAKE) term, which is composed of free particle (PAKE) and enthalpy
(EN) terms:
\begin{eqnarray}\label{MAKEs}
{T^{\rm MAKE}}^\mu_\nu &=& {T^{\rm MA}}^\mu_\nu - \rho_0 \uvec^\mu \etavec_\nu/\alpha ,\\
{T^{\rm PAKE}}^\mu_\nu &=& (\uvec_\nu - \etavec_\nu/\alpha)\rho_0 \uvec^\mu  \nonumber ,\\
{T^{\rm EN}}^\mu_\nu &=&  (e_g + p_g)\uvec^\mu \uvec_\nu + p_g \delta^\mu_\nu \nonumber ,
\end{eqnarray}
such that ${T^{\rm MAKE}}^\mu_\nu = {T^{\rm PAKE}}^\mu_\nu+{T^{\rm
    EN}}^\mu_\nu$.  Here, $e_g$ is the internal energy density and
$p_g=(\Gamma-1)e_g$ is the ideal gas pressure with adiabatic index
$\Gamma=4/3$ ($\Gamma=5/3$ may lead to somewhat different results ;
\citealt{mg04,mm07}).  The sound speed is given as $c_s=\sqrt{\Gamma
  p_g/(\rho_0 + e_g + p_g)}$.  The contravariant fluid-frame magnetic
4-field is given by $\bvec^\mu$, which is related to the lab-frame
3-field via $\bvec^\mu = \Bvec^\nu h^\mu_\nu/\uvec^t$ where $h^\mu_\nu
= \uvec^\mu \uvec_\nu + \delta^\mu_\nu$ is a projection tensor
that is the inverse metric on spatial hypersurfaces orthogonal to $u^\mu$
(giving an orthonormal comoving tetrad that is space-like diagonal)
and that extracts the spatial part of the comoving magnetic field
(giving $b^\mu u_\mu=0$)
and that in any basis gives $B^\mu = (-u^\nu B_\nu)u^\mu + h^\mu_\nu B^\nu$
so the field has been split into parts parallel and perpendicular to $u^\mu$,
and $\delta^\mu_\nu$ is the Kronecker delta function.  The magnetic energy
density ($u_b$) and pressure ($p_b$) are $u_b=p_b=\bvec^\mu
\bvec_\mu/2 = b^2/2$.  The total pressure is $p_{\rm tot} = p_g +
p_b$, and plasma $\beta_{\rm plasma}\equiv p_g/p_b$.  The 4-velocity of a zero
angular momentum observer (ZAMO) is $\etavec_\mu=\{-\alpha,0,0,0\}$
where $\alpha=1/\sqrt{-g^{tt}}$ is the lapse,
which is a convenient observer that follows worldlines with
$\dxvec^i/dt = -\betavec^i$ with $\betavec^i = \alpha^2 g^{ti}$
such that $\etavec^\mu = (1/\alpha)\{1,-\betavec^i\}$.
The BH metric can then be written as
$ds^2 = -\alpha^2 dt^2 + h_{ij}(\dxvec^i + \betavec^i dt)(\dxvec^j + \betavec^j dt)$,
which often simplifies calculations.
We used this ZAMO to construct a frame that moves relative
to the ZAMO with a time-like 4-velocity
of $\reluvec^\mu = \uvec^\mu - \gamma \etavec^\mu$ where
$\gamma=-\uvec^\alpha \etavec_\alpha$.  Spatial interpolations
of this relative 4-velocity always lead to time-like
4-velocities (unlike interpolations of the lab-frame 3-velocity)
and also give a unique time-like observer even inside the BH ergosphere
(unlike the lab-frame 4-velocity that can have two time-like observers inside the ergosphere).
So this relative 4-velocity is well-suited to numerical simulations
that rely heavily upon spatial interpolations and must not lose
any information about the fluid frame in the ergosphere.
For any 3-vector
(e.g. $\Bvec^i$), the ``quasi-orthonormal'' vector is $B_i\equiv
\Bvec^i \sqrt{g_{ii}}$ computed in spherical polar coordinates.

Magnetic flux conservation is given by the induction equation
\begin{equation}
\partial_t(\detg \Bvec^i) = -\partial_j[\detg(\Bvec^i \vvec^j - \Bvec^j \vvec^i)] ,
\end{equation}
where $g={\rm Det}(g_{\mu\nu})$ is the metric's determinant, and the
lab-frame 3-velocity is $\vvec^i = \uvec^i/\uvec^t$.  No explicit
viscosity or resistivity are included, but we used the energy
conserving HARM scheme so all dissipation is captured
\citep{gam03,mck06ffcode}.

The energy-momentum conservation equations are only modified due to
so-called numerical density floors that keep the numerical code stable
as described in detail in the Appendix of \citemtb.  The injected
densities were tracked and removed from all calculations.

\subsection{Diagnostics}
\label{sec:diagnostics}

The discussion in this section follows that of \citemtb.
Diagnostics were computed from snapshots produced every $\sim 2r_g/c$.
For quantities $Q$, averages over space ($\langle Q \rangle$) and time
($[Q]_t$) were performed directly on $Q$ (e.g. on $v_\phi$ rather than
on any intermediate values).  Any flux ratio vs. time with numerator
$F_N$ and denominator $F_D$ ($F_D$ often being mass or magnetic flux)
was computed as $R(t)=\langle F_N(t) \rangle / [\langle F_D
\rangle]_t$.  Time-averages were then computed as $[R]_t$.

\subsubsection{Fluxes and Averages}
\label{sec:avgradquant}

For flux density $F_d$, the flux integral is
\begin{equation}
F(r) \equiv \int dA_{23} F_d ,
\end{equation}
where $dA_{23}=\gdet d\xvec^{(2)} d\xvec^{(3)}$ ($dA_{\theta\phi}$ is
the spherical polar version).  For example, $F_d=\rho_0 \uvec^{(1)}$
gives $F=\dot{M}$, the rest-mass accretion rate.  For weight $w$, the
average of $Q$ is
\begin{equation}
Q_w(r) \equiv \langle Q \rangle_w \equiv \frac{\int dA_{\theta\phi} w Q}{\int dA_{\theta\phi} w} ,
\end{equation}
All $\theta,\phi$ angles are integrated over.

\subsubsection{Disk Thickness Measurement}
\label{sec:diskthick1}

The disk's geometric half-angular thickness is given by
\begin{equation}\label{thicknesseq}
H/R \equiv \left(\left\langle \left(\theta-\theta_0\right)^2 \right\rangle_{\rholab}\right)^{1/2} ,
\end{equation}
where we integrated over all $\theta$ for each $r,\phi$, and $\theta_0
\equiv \pi/2 + \left\langle \left(\theta-\pi/2\right)
\right\rangle_{\rholab}$ was also integrated over all $\theta$ for each
$r,\phi$, and the final $H/R(r)$ was from $\phi$-averaging with no
additional weight or $\detg$ factor.  This way of forming
$H/R(r)$ works for slightly tilted thin disks or disordered thick disks.
For a Gaussian distribution in density, this satisfies
$\rholab/(\rholab[\theta=0]) \sim \exp(-\theta^2/(2(H/R)^2))$.

The value of $H/R$ is computed at various locations and times,
while for this work we report the value at the initial pressure maximum in the untilted solutions
and at $r\sim 30r_g$ in the evolved quasi-steady state solutions.
The thickness at large distances from the BH
does not change significantly as the flow evolves \citemtb
or after tilt is introduced in the simulation.
Note that, unlike expected from modern thick disk accretion theory,
the BH jet compresses the disk causing the geometric $H/R$
to drop as the disk approaches the BH \citemtb.

\subsubsection{Relative Tilt Measurement}
\label{sec:tilt}

The tilt of the disk rotation axis and jet direction with respect to
the BH spin axis was measured using the angular
momentum\cite{2005ApJ...623..347F}.  The BH angular momentum vector is
\begin{equation}
J_{\rm BH} = \left( a M \sin\theta_{\rm tilt,0} \hat{x},~0,~a M \cos\theta_{\rm tilt,0} \hat{z} \right) ,
\end{equation}
with $\theta_{\rm tilt,0}$ is the initial tilt between the angular
momentum vectors of the BH and the disk.  The MHD angular momentum
vector is
\begin{equation}
J_{\rm MHD}(r) = \left[ (J_{\rm MHD})_x \hat{x},~(J_{\rm MHD})_y \hat{y},~(J_{\rm MHD})_z \hat{z} \right]
\end{equation}
in an asymptotically flat space-time, where
\begin{equation}
(J_{\rm MHD})_\rho = \frac{\epsilon_{\mu \nu \sigma \rho}
                         L^{\mu \nu} \zeta^\sigma}
                       {2}
\end{equation}
for some frame with 4-velocity $\zeta^\sigma$
(for simplicity, we choose the Kerr-Schild lab-frame), and
\begin{equation}
L^{\mu \nu} = \int \mathrm{d}\Sigma_\alpha \left(x^\mu T^{\nu \alpha} - x^\nu T^{\mu \alpha} \right) ,
\end{equation}
where $\mathrm{d}\Sigma_\alpha=\epsilon_{\beta\gamma\delta\alpha}
dx^\beta dy^\gamma dz^\delta$ and the integral is performed for each
radial shell.  The unit vector $\hat{y}$ points along the axis around
which the BH is tilted and the unit vector $\hat{z}$ points along the
initial rotational axis of the disk.

The plasma's relative tilt angle is\citenelson
\begin{equation}
\theta_{\rm tilt}(r) = \arccos\left[ \frac{J_{\rm BH} \cdot J_{\rm MHD}(r)}
{|J_{\rm BH}| |J_{\rm MHD}(r) |} \right] .
\end{equation}
Just after the restart from a
untilted simulation and instantly applying tilt, $\theta_{\rm
  tilt}(r)\approx \theta_{\rm tilt,0}$ throughout the disk.

As in \citemtb, the disk is defined with the spatial integral
performed with weight $\rho_0 u^t$ (using $\rho_0$ alone gives similar results)
and normalized by the weight's
volume integral.  The jet is defined by taking the integral only over
the region with $b^2/\rho_0\ge 1$ with no weight.

\subsubsection{Efficiency of Energy Extraction Measurement}
\label{fluxes}

The rest-mass flux, specific energy flux, and specific angular
momentum flux are respectively given by
\begin{eqnarray}\label{Dotsmej}
\dot{M} &=&  \left|\int\rho_0 \uvec^r dA_{\theta\phi}\right| , \\
\emath \equiv \frac{\dot{E}}{[\dot{M}]_t} &=& -\frac{\int T^r_t dA_{\theta\phi}}{[\dot{M}]_t} , \\
\jmath \equiv \frac{\dot{J}}{[\dot{M}]_t} &=& \frac{\int T^r_\phi dA_{\theta\phi}}{[\dot{M}]_t} .
\end{eqnarray}

The net flow efficiency (efficiency of energy extraction from the BH)
is given by
\begin{equation}\label{eff}
  \eff = \frac{\dot{E}-\dot{M}}{[\dot{M}]_t} = \frac{\dot{E}^{\rm EM}(r) + \dot{E}^{\rm MAKE}(r)}{[\dot{M_{\rm H}}]_t} . \\
\end{equation}
Positive values correspond to an extraction of positive energy from
the system at some radius.  Values greater than $100\%$ indicate that
more energy goes out of the BH than goes in.

\subsubsection{Inflow Equilibrium and Quasi-Steady State}
\label{sec_infloweq}

Inflow equilibrium is defined as when the flow is in a complete
quasi-steady-state and the accretion fluxes are constant (apart from
noise) vs. radius and time.  The inflow equilibrium timescale is
\begin{equation}\label{tieofrie}
t_{\rm ie} = N \int_{r_i}^{r_{\rm ie}} dr\left(\frac{-1}{[\langle v_r\rangle_{\rholab}]_t}\right) ,
\end{equation}
for $N$ inflow times from $r=r_{\rm ie}$ and $r_i=12r_g$ to focus on
the more self-similar part of the flow far from the BH.
The value of $r_{\rm ie}$ is measured directly from the simulation
using Equation~(\ref{tieofrie}) rather than indirectly from an estimate
of the $\alpha$ viscosity and disk scale-height.
As in our prior work \citemtb, we define the quasi-steady state
out to some radius to mean that inflow equilibrium has been reached
out to that radius under the condition that $N$ is between $2$ and $3$.

\subsubsection{Modes and Correlation Lengths}
\label{sec_cor}

The flow structure was studied via the discrete Fourier transform of
$dq$ (related to quantity $Q$) along $x=r,\theta,\phi$ giving
amplitude $a_p$ for $p=n,l,m$, respectively.  The averaged amplitude
is
\begin{equation}\label{am}
\left\langle \left|a_p\right| \right\rangle \equiv \left\langle \left|\mathcal{F}_p\left(dq\right)\right| \right\rangle \equiv \int_{\rm not\ x} \left| \sum_{k=0}^{N-1}dq~{\rm e}^{\frac{-2\pi i p k}{N-1} } \right| ,
\end{equation}
computed at $r=r_{\rm H},4r_g,8r_g,30r_g$.  The $x$ is one of
$r,\theta,\phi$ and ``not x'' are others (e.g. $\theta,\phi$ for
$x=r$).  The $dq$ is (generally) a function of $x$ on a uniform grid
indexed by $k$ of $N$ cells that span: $\delta r$ equal to $0.75r$
around $r$ for $x=r$, $\pi$ for $x=\theta$, and $2\pi$ for
$x=\phi$. The $N$ is chosen so all structure from the original grid is
resolved, while the span covered allows many modes to be resolved.

For all $x$, $dq \equiv \detg d\xvec^{(1)} d\xvec^{(2)} d\xvec^{(3)}
\delta Q/q_N$.  For $x=r,\theta$, we let $q_N=\int_{\rm\ not\ x} \detg
d\xvec^{(1)} d\xvec^{(2)} d\xvec^{(3)} \langle [Q]_t \rangle$,
$\langle [Q]_t \rangle$ as the time-$\phi$ averaged $Q$, and $\delta Q
= Q - \langle[Q]_t\rangle$.  Using $dq$ removes gradients with
$r,\theta$ so the Fourier transform acts on something closer to
periodic with constant amplitude (see also \citealt{bas11}).  For
$x=\phi$, we let $q_N=1$ and $\delta Q = Q$ because the equations of
motion are $\phi$-ignorable.  For $x=\theta,\phi$, the radial integral
is computed within $\pm 0.1r$.  For $x=r,\theta$, the $\phi$ integral
is over all $2\pi$.  For $x=r,\phi$, the $\theta$ integral is over all
$\pi$.  For all $x$ cases, the $\theta$ range of values is over
either the disk or jet (as defined earlier),
where these regions are defined via $\phi$-averaged
quantities at each time. Notice we averaged the mode's absolute
amplitude, because the amplitude of $\langle \delta Q\rangle$
de-resolves power (e.g. $m=1$ out of phase at different $\theta$ gives
$\langle \delta Q\rangle \to 0$ and $a_m\to 0$) and is found to
underestimate small-scale structure.

We also computed the correlation length: $\lambda_{x,\rm cor}=x_{\rm
  cor}-x_0$, where $x_0=0$ for $x=\theta,\phi$ and $x_0$ is the inner
radius of the above given radial span for $x=r$, where $n_{\rm cor} =
\delta r/\lambda_{r,\rm cor}$, $l_{\rm cor}=\pi/\lambda_{\theta,\rm
  cor}$, and $m_{\rm cor} =(2\pi)/\lambda_{\phi,\rm cor}$.  The
Wiener-Khinchin theorem for the auto-correlation gives
\begin{equation}\label{mcor}
\exp(-1) = \frac{\mathcal{F}^{-1}_{x=x_{\rm cor}}[\langle|a_{p>0}|\rangle^2]}{\mathcal{F}^{-1}_{x=x_0}[|\langle a_{p>0}|\rangle^2]} ,
\end{equation}
where $\mathcal{F}^{-1}[\langle |a_{p>0}|\rangle ^2]$ is the inverse
discrete Fourier transform of $\langle |a_p|\rangle^2$ but with
$\langle a_0\rangle$ reset to $0$ (i.e. mean value is excluded).

Turbulence (or flow structure in general) is resolved for grid cells per correlation length,
\begin{equation}\label{Qmcor}
Q_{p,\rm{}cor} \equiv \frac{\lambda_{x,\rm cor}}{\Delta_{x}} ,
\end{equation}
of $Q_{p,\rm{}cor}\ge 6$, for $x=r,\theta,\phi$ and $p=n,l,m$,
respectively.  Otherwise, modes are numerically damped on a dynamical
timescale (even $Q=5$ would not indicate the mode is marginally
resolved, because numerical noise can keep $Q\approx 5$ at increasing
resolution until finally the mode is actually resolved -- finally
leading to an increasing $Q\ge 6$ with increasing resolution ; as seen
by \citealt{sdgn11}).  Reported $Q_{p,\rm{}cor}$ take $1/\Delta_{x}$
as the number of grid cells covering the span of $\lambda_{x,\rm cor}$
as centered on: middle of $x^{(1)}$ within the used radial span for $x=r$, $\theta\sim \pi/2$
(a reasonable approximation even for tilted models when this is measured at $r\approx 8r_g$)
for $x=\theta$ for the disk and $\theta\sim 0$ for $x=\theta$ for the
jet, and anywhere for $x=\phi$.

\subsubsection{The Magneto-Rotational Instability}
\label{sec_mri}

The MRI is a linear instability with fastest growing wavelength of
\begin{equation}\label{lambdamri}
\lambda_{x,\rm MRI} \approx  2\pi \frac{|v_{x,\rm A}|}{|\Omega_{\rm rot}|} , \\
\end{equation}
for $x=\theta,\phi$, where $|v_{x,\rm A}|=\sqrt{\bvec_x
  \bvec^x/\epsilon}$ is the $x$-directed~\alf~speed, $\epsilon\equiv
b^2 + \rho_0 + e_g + p_g$, and $r\Omega_{\rm rot} = v_{\rm rot}$.
$\lambda_{\rm MRI}$ is accurate for $\Omega_{\rm rot}\propto r^{-5/2}$
to $r^{-1}$.  $\Omega_{\rm rot},v_{\rm A}$ were separately
angle-volume-averaged at each $r,t$.

The MRI is resolved for grid cells per wavelength
in the $x=\theta,\phi$ directions,
\begin{equation}\label{q1mri}
\Qx \equiv \frac{\lambda_{x,\rm MRI}}{\Delta_{x}} ,
\end{equation}
of $Q_{\theta}\gtrsim 10$ and $Q_{\rm \phi}\gtrsim 20$ \cite{hgk11},
where $\Delta_{r} \approx
d\xvec^{(1)} (dr/d\xvec^{(1)})$, $\Delta_{\theta} \approx r
d\xvec^{(2)} (d\theta/d\xvec^{(2)})$, and $\Delta_{\phi} \approx r
\sin\theta d\xvec^{(3)} (d\phi/d\xvec^{(3)})$.  Volume-averaging was applied to $v_{x,\rm A}/\Delta_{x}$ and
$|\Omega_{\rm rot}|$ were separately $\theta,\phi$-volume-averaged
before forming $\Qx$.

The MRI suppression factor corresponds to the number of MRI
wavelengths across the full disk:
\begin{equation}\label{q2mri}
\Qtwo \equiv \frac{2 r (H/R)}{\lambda_{\theta,\rm MRI}} .
\end{equation}
Wavelengths $\lambda<0.5\lambda_{\theta,\rm MRI}$ are stable, so the
linear MRI is suppressed for $\Qtwo<1/2$ when no unstable wavelengths
fit within the full disk \citep{bh98,pp05}.  $\Qtwo$ (or $\Qtwoweak$)
uses averaging weight $w=(b^2\rholab)^{1/2}$ (or $w=\rholab$),
condition $\beta_{\rm plasma}>1$, and excludes regions where density floors are
activated. Weight $w=(b^2\rholab)^{1/2}$ is preferred, because much
mass flows in current sheets where the magnetic field vanishes and yet
the MRI is irrelevant.  When computing the averaged $\Qtwo$, $v_{\rm
  A}$ and $|\Omega_{\rm rot}|$ were separately
$\theta,\phi$-volume-averaged within $\pm 0.2r$ for each $t,r$.  The
averaged $\Qtwo$ is at most $30\%$ smaller than $\Qtwoweak$.

A measure of the magnetic stress per unit pressure given by
\begin{equation}\label{alphaeq}
\alpha_{\rm mag} \approx -\frac{b_r (\bvec_\phi \sqrt{g^{\phi\phi}})}{p_b} ,
\end{equation}
can be used to determine how well-converged a numerical solution might
be for MRI-dominated disk simulations.  A value of $\alpha_{\rm
  mag}\ge 0.4$ indicates a well-converged MRI-dominated simulation
\cite{hgk11,2011arXiv1106.4019S}.  The numerator and denominator are
separately volume averaged in $\theta,\phi$ for each $r$.  Note that
$\sin(2\theta_b)=\alpha_{\rm mag}$ for tilt angle $\theta_b$
\citep{2011arXiv1106.4019S}.
This measure is weakly applicable for our solutions where
the MRI is suppressed and strong effects occur due to BH spin.
However, at least for our thinner disk models this measure
seems to roughly be applicable.  For our thick disk models,
the large amount of trapped magnetic flux and high BH spins lead to this
measuring more about the angular momentum transported out
from the BH into the disk than the MRI within the disk.
For such thick disks, retrograde spins lead to negative
$\alpha_{\rm mag}$ \citemtb.

\subsection{Physical and Numerical Models}
\label{sec:physnummodels}

This section describes our models.  The model names are in the form
AxByNzTp, where x is the approximate value of the BH spin, y
identifies the field geometry (p=poloidal, f=flipping poloidal,
t=toroidal), z identifies the normalization of the magnetic field, and
p identifies the tilt (tilt is zero if not given).  This is the same
type of notation we have used previously\citemtb, where for this
paper only poloidal or poloidal flipping models are considered. For
instance, our model A0.94BfN40T0.3 had a spinning BH ($j=0.9375$), a
poloidal field that flips polarity with radius, $\beta_{\rm plasma,
min}\approx 40$ (i.e. smallest value of plasma $\beta_{\rm plasma}$ is
$40$), and a tilt of $0.3$ radians.

All tilted models are continuations of the models without tilt given
elsewhere\citemtb with parameters given in their tables~1 and~2.
The relative tilt was imposed (via a change in the BH spin axis) at a
time $t=8000r_g/c$ for thick disk models, $t=10000r_g/c$ for thinner
disk models except for the $a=0.99$ models that used $t=15000r_g/c$.
The thick disk models ran from $8000r_g/c$ to roughly $17000r_g/c$,
after which the initial magnetic field polarity inversion reaches
the black hole, so no measurements are reported after this time.
The thinner disk models with $|j|=0.9$ ran from $10000r_g/c$ to
roughly $22000r_g/c$, and the thinner disk models with $j=0.99$ ran
from $15000r_g/c$ to roughly $27000r_g/c$.  These thinner disk
models have no polarity inversions in the initial conditions.
These durations allow the
disk and jet to reach a quasi-steady state out to about $r\sim 40r_g$
within which results are reported.

\subsubsection{Physical Models}
\label{sec:modelsetup}

The details of the original simulation's initial mass distribution (an
isentropic hydro-equilibrium torus), magnetic field distribution
(poloidal magnetic field loops), and velocity are as provided in
\citemtb.  We only studied magnetic field configurations that led
to a saturated magnetic field strength value near the BH, and we only
studied cases that generate sufficiently large-scale magnetic field to
produce a jet \cite{bhk08,mb09}.

Our models were designed so
that near the horizon the poloidal ($R-z$) magnetic flux reached a
saturation point independent of the initial poloidal magnetic flux
\citemtbandsasha, which allowed
us to provide robust results that are
insensitive to any larger amount of magnetic flux available in the
surrounding medium \cite{nia03}.
In this sense, we identify the magnetic field strength reached as ``natural''
because it would require fine-tuning or an uncommon accident
for the magnetic flux supply
to be near saturation but somehow not simply
far below or far beyond that required for near-BH saturation.
A supply far below saturation would not lead to powerful jets as required
by observations for many systems,
while a large supply would simply lead to saturation
out to some arbitrarily large distance and so not affect the near-BH dynamics.

Our initial magnetic flux supply was still finite,
and our $H/R\sim 0.3$ models did not (and would not) reach a fully saturated
state beyond $r\sim 30r_g$ \citeretro,
while our $H/R\sim 0.6$ models
have enough magnetic flux supply
to reach magnetic flux saturation out to $r\sim 100r_g$ \citemtb.

Our general relativistic magnetized disk simulations started with a
tilted BH whose spin originally pointed towards the
vertical ($z$) direction \citemtb, and then (for this work) that
BH spin direction was suddenly rotated by an angle $\theta_{\rm tilt,0}$
around another perpendicular axis ($y$-axis for definiteness) leading
to the spin axis pointing more towards the $x$-axis for larger
$\theta_{\rm tilt,0}$.  A value of $\theta_{\rm tilt,0}=0$ means no tilt,
while $\theta_{\rm tilt,0}=90^\circ$ means maximal tilt, while
$\theta_{\rm tilt,0}=180^\circ$ means the BH was rotated so much that the
spin value of $j$ is simply the opposite sign.  Hence, by symmetry
we only need to consider $\theta_{\rm tilt,0}=0^\circ$ to $90^\circ$ for
dimensionless spins in the range $-1\le j \le 1$.

For thick disks with height ($H$) to radius ($R$) ratio of $H/R\approx
0.6\approx 34^\circ$ (the evolved quasi-steady state value of $H/R$ at $r\sim 30r_g$,
where $H/R\approx 0.6$ at the initial torus pressure maximum at $r=100r_g$),
we obtained quasi-steady solutions with tilts
$\theta_{\rm tilt,0}=0^\circ,20^\circ,40^\circ$, $90^\circ$ with spin
$j=+0.9375$.  Note that $j=-0.9375$ is expected to behave
similarly to a model with $j=+0.9375$ because these thick disk
models result in the entire disk being forced to rotate in the same
direction as the BH spin -- giving a result that is completely
independent of the initial disk angular momentum out to $r\sim 40r_g$
\citemtb.

For thinner disks with $H/R\approx 0.3\approx 17^\circ$
(the evolved quasi-steady state value of $H/R$ at $r\sim 30r_g$,
where $H/R\approx 0.2$ at the initial torus pressure maximum at $r\sim 35r_g$),
we obtained quasi-steady solutions with tilts $\theta_{\rm
  tilt,0}=0^\circ,9^\circ,17^\circ,34^\circ$,$90^\circ$ with spins
$j=-0.9,+0.9,+0.99$.

Note that $j$ is really a proxy for the actual BH angular frequency
given by $\Omega_{\rm H}= jc/2r_{\rm H}$ (maximum value of $0.5$) that
sets the rate of frame-dragging.  A value of
$j=0.5,0.9,0.9375,0.99,1$ gives $\Omega_{\rm H}\approx
0.13,0.31,0.35,0.43,0.5$.  Hence, a value of $j=0.9$ is really a
middle-range value for the BH rotation rate.  For small values of
$j\lesssim 0.5$ that operate at less than a fifth of the maximum
rotation rate, the effects of frame-dragging or tilt are expected to
be none to minimal.  Of course, in the limit of $j=0$ then the BH
space-time is spherical and there is no physical manifestation of any
chosen $\theta_{\rm tilt,0}$.

Table~S1 gives a list of diagnostics that we have measured as
discussed in section~\ref{sec:diagnostics}.  These diagnostics
are measured just like in our prior work on untilted simulations \citemtb.
A list of identical diagnostic measurements for other untilted
simulations (including those used as a starting point for the tilted
simulations) can be found in extensive tables elsewhere \citemtb.
The table shows that $\Upsilon$ and $\eta$ remain relatively high for all tilted
models, with some drop seen for fully tilted models.

\subsubsection{Numerical Models}
\label{sec:numsetup}

The resolution is chosen equal or higher than required to resolve
turbulence and physical disk instabilities, as discussed previously
\citemtb.  We chose a resolution
$N_r\times N_\theta\times N_\phi$ that had a grid aspect ratio of
1:1:1 for most of the domain for $r\lesssim 40r_g$.  The grid is fully
3D with a $\phi$-range of $2\pi$.  This allowed the $\phi$ dimension to
be treated equally to the $r-\theta$ dimensions.  Thick disk models
used a resolution of $N_r\times N_\theta\times N_\phi=272\times 128
\times 256$, while thinner disk models used $N_r\times N_\theta\times
N_\phi=288\times 128\times 128$ (twice larger $N_\phi$ than our prior
models for $|j|=0.9$).

In our prior work on otherwise identical untilted simulations
\citemtb, we used numerous diagnostics and numerical convergence
quality measures to check that structures in the disk and jet were
properly resolved (our measures are extensions of those discussed
elsewhere
\cite{2009ApJ...694.1010G,hgk11,sdgn11,bas11,2011arXiv1106.4019S}).
In our prior study, for our thick disk models, we found that
$272\times 128\times 256$ was sufficient to overly resolve the MRI,
well resolve the radial and azimuthal correlation lengths that
describe the characteristic sizes of structures in the disk and jet,
and marginally resolve the vertical correlation length due to magnetic
Rayleigh-Taylor modes that lead to disk compression.  All thinner disk
models were similarly resolved, except the $|j|=0.9$ models that used
$N_\phi=64$ did slightly under-resolve the azimuthal correlation
length by a factor less than two.  In convergence studies in that
paper, the number of grid cells per azimuthal correlation length was
seen to scale with resolution as expected, so in this paper we doubled
the $N_\phi$ resolution (from $N_\phi=64$ to $N_\phi=128$) to ensure
that untilted thinner disk $|j|=0.9$ simulations would properly
resolve all correlation lengths.

For these new tilted simulations, as in our prior work \citemtb,
we confirmed that the MRI and turbulence were resolved both initially
and at late times through-out the disk and jet at many different
radii. Table~S2 gives a list of convergence quality measures
as computed in section~\ref{sec:diagnostics}.
These measures were computed the same way as in our prior work \citemtb.
A list of identical convergence quality
measurements for the untilted simulations (used as a starting point
for the tilted simulations) can be found in extensive tables elsewhere
\citemtb.  Since the tilted simulations just continue the untilted
simulations, the initial MRI convergence quality measures were the same as already
reported elsewhere \citemtb.  In all the untilted models that we
used for the tilted simulations, the initial MRI was well-resolved.

In all cases, the convergence quality measures indicated that the
simulations are at least marginally converged and some aspects were
well-converged, which is expected given that was how we chose the
resolutions as based upon similar measures for otherwise identical
untilted simulations.

As in our prior work \citemtb, we also explicitly checked
convergence of all our measures, including for disk tilt and jet tilt at late
times, by repeating our thinner disk $j=0.99$ $\theta_{\rm
tilt}=1.5708$ simulations with a resolution of $144\times 64\times 32$
and $144\times 64\times 64$.  We did not double the resolution to
$576\times 256\times 256$ because such a single model would use about
16 million central processing unit (CPU)-hours on Kraken.  Further, as we see next, convergence was
already been obtained for $\Upsilon$ and $\theta_{\rm tilt}$ that we
are focused on.

As found in our prior work for untilted models, the value of
$\Upsilon$ that controls the jet magnetic flux was insensitive to such
changes in resolution.  The disk and jet align with $\theta_{\rm
rot}\approx 0.1$ at $r\approx 4r_g$ and $\theta_{\rm rot}\approx 0.6$
at $r\approx 30r_g$ as with the higher resolution model.  The jet
magnetic flux is what leads to the EM force that aligns the disk with
the BH, so the ``magneto-spin alignment'' effect is not sensitive to
such changes in resolution.  This is expected because the amount of
trapped magnetic flux is set primarily by non-turbulent force balance
against the large-scale magnetic field, and the EM jet power (and so
force) does not strongly depend upon resolving the small-scale
turbulence in the disk.

The entire set of new tilted simulations (as continuations of untilted
simulations) required about 22 million CPU-hours on the
Extreme Science and Engineering Discovery Environment (XSEDE) Kraken
cluster and another 11 million CPU-hours on the NAS Pleiades Westmere
cluster.

\newpage

\begin{figure*}
\centering
\includegraphics[width=4.0in,clip]{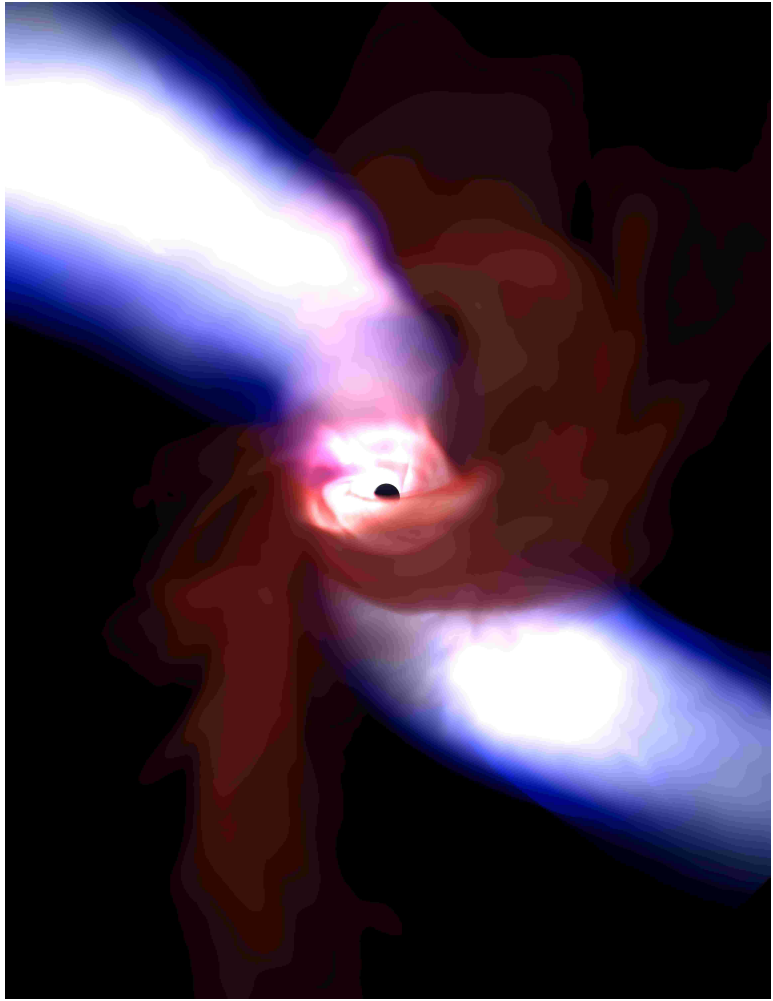}
\caption{Artistic 3D rendering showing an
  evolved 3D snapshot, similar to Fig.~1 in the report,
  for a model with $j=0.99$, initial relative tilt
  $\theta_{\rm tilt,0}\approx90^\circ$, and disk thickness of
  $H/R\sim 0.3$.  The rotating BH is shown as a black sphere and
  sits at the center of the box of size of about $r=-40r_g$ to
  $r=+40r_g$ in the vertical dimension and otherwise about $r=-30r_g$
  to $r=+30r_g$ in the other dimensions.  The jet is shown by the
  region where the magnetic energy density per unit rest-mass energy
  density is greater than about $10$ up to about $100$ (blue-white
  volume rendering).  The logarithm of the density in the disk region
  is shown close to the BH (red-white volume rendering).  The BH spin
  axis points vertically.  Near the BH, the disk rotational axis and
  jet direction are strongly aligned with the BH spin axis.  However,
  at $t=0$ and at large radius for all evolved times, the disk and jet
  axes pointed completely horizontally, i.e. perpendicular to the
  vertical axis.  This shows the strong magnetic torques have caused
  the disk and jet to change orientation by $90^\circ$ and to become
  aligned with the BH spin axis near the BH.  At larger distances, the
  jet gradually bends away from the BH spin axis and aligns with the
  disk rotational axis at large distances.  This leads to the jet and
  counter-jet being offset at large distances.}
\label{cover}
\end{figure*}

\newpage

\begin{table*}\footnotesize
\caption{Simulation Physical Setup and Results}
\begin{center}
\begin{tabular}[h]{|l|r|r|r|r|r|r|r|r|r||r|}
\hline
ModelName         &   $j$    & $\theta_{\rm tilt,0}$   &  $T_f$  &  $T^a_i$--$T^a_f$  &   $r_{\rm{}in}$ &   $r_{\rm{}out}$ &  $H/R$  & $\dot{M}_{\rm{}H}$   &  $\Upsilon_{\rm{}H}$  &   $\eta_{\rm{}H}$  \\
\hline
A0.94BfN40        &  0.9375  &  0                    &  26548  &   8000--17000      &  12             &  29              &  0.6    &  48                  &  17                   &  240            \\  
A0.94BfN40T0.35   &  0.9375  &  0.35                 &  20376  &  14000--17000      &  12             &  30              &  0.59   &  42                  &  17                   &  250            \\  
A0.94BfN40T0.7    &  0.9375  &  0.7                  &  18656  &  14000--17000      &  12             &  30              &  0.59   &  67                  &  12                   &  67             \\  
A0.94BfN40T1.5708 &  0.9375  &  1.5708               &  18132  &  14000--17000      &  12             &  30              &  0.68   &  43                  &  12                   &  73             \\  
A-0.9N100         &  -0.9    &  0                    &  20060  &  11000--20060      &  12             &  30              &  0.31   &  18                  &  6.6                  &  32             \\  
A-0.9N100T0.15    &  -0.9    &  0.15                 &  23280  &  17000--23280      &  12             &  30              &  0.35   &  21                  &  5.4                  &  18             \\  
A-0.9N100T0.3     &  -0.9    &  0.3                  &  22095  &  17000--22095      &  12             &  30              &  0.36   &  23                  &  4.5                  &  10             \\  
A-0.9N100T0.6     &  -0.9    &  0.6                  &  20400  &  17000--20400      &  12             &  30              &  0.35   &  22                  &  3.7                  &  7              \\  
A-0.9N100T1.5708  &  -0.9    &  1.5708               &  23580  &  17000--23580      &  12             &  30              &  0.37   &  22                  &  4.2                  &  11             \\  
A0.9N100          &  0.9     &  0                    &  19895  &  11000--19895      &  12             &  30              &  0.31   &  11                  &  10                   &  100            \\  
A0.9N100T0.15     &  0.9     &  0.15                 &  20290  &  17000--20290      &  12             &  30              &  0.32   &  6.1                 &  12                   &  150            \\  
A0.9N100T0.3      &  0.9     &  0.3                  &  22585  &  17000--22585      &  12             &  30              &  0.35   &  14                  &  8.8                  &  69             \\  
A0.9N100T0.6      &  0.9     &  0.6                  &  21010  &  17000--21010      &  12             &  30              &  0.34   &  14                  &  8.9                  &  70             \\  
A0.9N100T1.5708   &  0.9     &  1.5708               &  24385  &  17000--21000      &  12             &  30              &  0.41   &  22                  &  4.8                  &  16             \\  
A0.99N100         &  0.99    &  0                    &  31400  &  15000--31400      &  12             &  30              &  0.33   &  9.8                 &  9.1                  &  130            \\  
A0.99N100T0.15    &  0.99    &  0.15                 &  25120  &  22000--25120      &  12             &  30              &  0.36   &  11                  &  9                    &  130            \\  
A0.99N100T0.3     &  0.99    &  0.3                  &  25085  &  22000--25085      &  12             &  30              &  0.36   &  9.3                 &  8.1                  &  110            \\  
A0.99N100T0.6     &  0.99    &  0.6                  &  25965  &  22000--25965      &  12             &  30              &  0.38   &  7.8                 &  9.3                  &  140            \\  
A0.99N100T1.5708  &  0.99    &  1.5708               &  28690  &  22000--28690      &  12             &  30              &  0.55   &  11                  &  5.2                  &  34             \\  
\hline
\hline
\end{tabular}
\end{center}
\caption{\addtocounter{table}{-1}The columns correspond to (1) the model name, (2)
BH spin, (3) initial tilt, (4) final time of the simulation, (5)
time-averaging range for any time-averaged quantity, (6) inner radius
used to measure some quantities (${\Qx}_i,{\Qtwo}_i$ in
Table~S2), (7) outer radius used to measure some quantities
(${\Qx}_o,{\Qtwo}_o$ in Table~S2), (8) accurate disk
thickness at $r\approx 30r_g$, (9) mass accretion rate, (10)
dimensionless magnetic flux on the horizon, and (11) BH efficiency.}
\label{tbls1}
\end{table*}

\newpage

\addtocounter{table}{+1}

\begin{table*}
\caption{Simulation Convergence Study}
\begin{center}
\begin{tabular}[h]{|l|r|r|r|r|r|r|r|r|r|r|}
\hline
ModelName         &  $\alpha_{\rm{}mag}$    &   $\bfrac{Q_{n,\rm{}cor,}}{{}_{\{\rho_0,b^2\}}}$     &    $\bfrac{Q_{l,\rm{}cor,}}{{}_{\{\rho_0,b^2\}}}$                                 &    $\bfrac{Q_{m,\rm{}cor,}}{{}_{\{\rho_0,b^2\}}}$  &   $\bfrac{Q_{\theta,\rm{}MRI}}{{}_{\{i,  o\}}}$ &   $\bfrac{Q_{\phi,\rm{}MRI}}{{}_{\{i,  o\}}}$   &    $\bfrac{S_{\rm{}d,\rm{}MRI}}{{}_{\{i,  o\}}}$  \\
\hline
A0.94BfN40        &  0.26                   &  22,                                             21  &                                                6,  12                                              &  19,                       18     &  110,                    120    &  380,                      540    &   0.33,  0.34  \\  
A0.94BfN40T0.35   &  0.22                   &  26,                                             25  &                                               10,  14                                              &  36,                       35     &  110,                    120    &  460,                      710    &   0.33,  0.34  \\  
A0.94BfN40T0.7    &  0.35                   &  27,                                             27  &                                               18,  20                                              &  31,                       32     &  160,                    130    &  440,                      550    &   0.24,  0.35  \\  
A0.94BfN40T1.5708 &  0.32                   &  25,                                             27  &                                               14,  17                                              &  23,                       21     &  140,                    130    &  330,                      610    &   0.24,  0.33  \\  
A-0.9N100         &  0.49                   &  23,                                             22  &                                                9,  11                                              &  10,                       10     &  110,                    35     &  44,                       42     &   0.29,  0.84  \\  
A-0.9N100T0.15    &  0.51                   &  20,                                             21  &                                               10,  15                                              &  8,                        10     &  100,                    77     &  42,                       54     &   0.4,   0.42  \\  
A-0.9N100T0.3     &  0.52                   &  22,                                             21  &                                               15,  20                                              &  9,                        13     &  70,                     71     &  39,                       54     &   0.55,  0.48  \\  
A-0.9N100T0.6     &  0.47                   &  22,                                             22  &                                               16,  20                                              &  9,                        9      &  99,                     62     &  54,                       57     &   0.39,  0.55  \\  
A-0.9N100T1.5708  &  0.45                   &  22,                                             22  &                                               24,  28                                              &  11,                       14     &  97,                     83     &  51,                       74     &   0.41,  0.43  \\  
A0.9N100          &  0.54                   &  29,                                             28  &                                               8,   10                                              &  8,                        8      &  120,                    42     &  60,                       44     &   0.27,  0.79  \\  
A0.9N100T0.15     &  0.48                   &  27,                                             26  &                                               9,   13                                              &  8,                        8      &  160,                    56     &  69,                       50     &   0.24,  0.59  \\  
A0.9N100T0.3      &  0.53                   &  27,                                             27  &                                               12,  17                                              &  8,                        13     &  100,                    69     &  62,                       56     &   0.38,  0.46  \\  
A0.9N100T0.6      &  0.49                   &  27,                                             28  &                                               17,  21                                              &  8,                        15     &  87,                     60     &  55,                       48     &   0.41,  0.56  \\  
A0.9N100T1.5708   &  0.38                   &  30,                                             29  &                                               24,  28                                              &  9,                        12     &  100,                    68     &  79,                       54     &   0.45,  0.55  \\  
A0.99N100         &  0.53                   &  22,                                             20  &                                               6,   8                                               &  7,                        7      &  110,                    67     &  66,                       54     &   0.33,  0.56  \\  
A0.99N100T0.15    &  0.5                    &  22,                                             20  &                                               6,   7                                               &  7,                        8      &  93,                     87     &  61,                       59     &   0.41,  0.41  \\  
A0.99N100T0.3     &  0.5                    &  22,                                             20  &                                               7,   7                                               &  7,                        8      &  130,                    80     &  72,                       55     &   0.3,   0.47  \\  
A0.99N100T0.6     &  0.46                   &  20,                                             20  &                                               6,   7                                               &  8,                        10     &  75,                     53     &  65,                       58     &   0.49,  0.64  \\  
A0.99N100T1.5708  &  0.24                   &  22,                                             23  &                                               11,  11                                              &  12,                       13     &  100,                    120    &  86,                       96     &   0.42   0.35  \\  
\hline
\hline
\end{tabular}
\end{center}
\caption{\addtocounter{table}{-1}The columns correspond to (1) the model name, (2)
$\alpha_{\rm mag}$ with radial averaging over $r=12r_g$ to $r=20r_g$,
(3) number of grid cells per correlation length in the radial ($n$)
direction, (4) number of grid cells per correlation length in the
vertical-angular ($l$) direction, (5) number of grid cells per
correlation length in the azimuthal ($m$) direction (all correlation
lengths are measured for the rest-mass density and magnetic energy
density at $r\approx 8r_g$, where beyond the horizon all values are
similar), (6) number of grid cells per vertical-angular MRI wavelength,
(7) number of grid cells per azimuthal MRI wavelength, and (8) MRI
suppression factor.  The inner ($i$) and outer ($o$) radius values for
columns 6,7,8 are measured at the inner and outer radii given in
Table~S1.}
\label{tbls2}
\end{table*}

\newpage

\section*{Movie S1}
\label{sec:movies1}

This movie shows inner distances for the evolved state of the model
with $j=0.99$, $\theta_{\rm tilt,0}=1.5708$, and $H/R\sim 0.3$ as
shown in Figs. 1-2 in the report (See also http://youtu.be/XxIzazsE-sg
,20MB mp4 for full 1080p HD).  The movie shows the full 3D structure
that is difficult to see in a single 2D image because the BH, jet, and
disk have axes that do not sit in the same two-dimensional plane.  The
movie shows the same rendering as Fig.~1 (without the vector field
surface and with density volume rendering replaced by density
isosurface) with the view rotated around a single axis so one can
clearly see the full 3D structure of the BH spin direction, disk plane
near the BH, jet direction, and disk rotational axis at larger
distances (as represented by the orange cylinder).

\section*{Movie S2}
\label{sec:movies2}

This movie shows the outer distances for the evolved state of the
model with $j=0.99$, $\theta_{\rm tilt,0}=1.5708$, and $H/R\sim
0.3$ as shown in Figs. 1-2 in the report (See also
http://youtu.be/25WGtFVcHe0 ,11MB mp4 for full 1080p HD).  The movie
shows the same rendering as Fig.~2 with the view rotated around a
single axis so one can clearly see the full 3D structure of the jet
near the BH, the disk plane near the BH, the jet direction as it
deviates at larger radii, the disk material that is pushed back into a
warp by the jet, and the disk rotational axis at larger distances (as
seen in the outer disk isosurface and represented by the orange
cylinder).

\newpage

\bibliography{bib}

\bibliographystyle{Science}